\documentclass[11pt,tightenlines,nofootinbib,superscriptaddress]{revtex4}
\usepackage{amsfonts,amssymb,amsthm,bbm}

\usepackage{amssymb,amstext,amsmath,amsthm}
\usepackage{MnSymbol}
\usepackage[dvips]{graphicx}
\usepackage{latexsym}
\usepackage{psfrag}
\usepackage{amsfonts}
\usepackage{bm} % bold math
\usepackage{bbm}
\usepackage{color}

\newcommand{\be}{\begin{equation}}
\newcommand{\ee}{\end{equation}}
\newcommand{\barray}{\begin{array}}
\newcommand{\earray}{\end{array}}
\newcommand{\bea}{\begin{eqnarray}}
\newcommand{\eea}{\end{eqnarray}}
\newcommand{\bs}{\begin{subequations}}
\newcommand{\es}{\end{subequations}}
\newcommand{\balign}{\begin{align}}
\newcommand{\ealign}{\end{align}}
\newcommand{\equ}{\begin{equation}}
\newcommand{\nequ}{\end{equation}}
\newcommand{\eqa}{\begin{eqnarray}}
\newcommand{\neqa}{\end{eqnarray}}

\def\nn{\nonumber}
\newcommand{\Ref}[1]{(\ref{#1})}

\newcommand{\bra}[1]{\la {#1}|}
\newcommand{\ket}[1]{|{#1}\ra}

\newcommand{\mrm}[1]{\quad \mathrm{#1}\quad}

\def\Ga{\Gamma}

\def\d{\delta}

\def\a{\alpha}
\def\b{\beta}

\def\e{\epsilon}
\def\bphi{\bm\varphi}
\def\brd{\bm\rd}
\def\bj{\bm j}
\def\bpi{\bar{\pi}}

\newcommand{\rd}{\mathrm{d}}
\newcommand{\p}{\partial}
\newcommand{\N}{\nabla}
\def\cL{ {\cal L}}
\def\i{\imath}

\def\bra{\langle}
\def\ket{\rangle}

\def\f{\frac}
\def\i{\imath}
\def\ot{{t^{\perp}}}

\def\bS{\bar{\Sigma}}

\def\bs{\bar{s}}
\def\bn{\bar{n}}
\def\bbn{\bm{\bar{n}}}
\def\bbs{\bm{\bar{s}}}
\def\bhr{\bm{\hat{r}}}
\def\aa{\bm{a}}

\newcommand{\un}[1]{\underline{\bm{#1}}}

\def\n{\bm{n}}
\def\t{\bm{t}}

\def\s{\bm{s}}
\def\r{\bm{r}}

\def\bht{\bm{\hat{t}}}
\def\bot{\bm{t^{\perp}}}
\def\bxi{\bm{\xi}}
\def\dd{\!\cdot \!}
\def\bmo{\bm{\omega}}

\DeclareMathOperator{\tr}{Tr}

\usepackage{bbm}

\begin{document}

\title{ Gravitational Energy, Local Holography and  Non-equilibrium Thermodynamics}
\author{Laurent Freidel}
\smallskip 
\affiliation{Perimeter Institute for Theoretical Physics, 31 Caroline St, N, On N2L 2Y5,
Waterloo, Canada} 
%\date{\today}
\begin{abstract}
We study the properties of gravitational system in finite regions bounded by gravitational screens.
We present the detail construction of the total energy of such regions and of the energy and momentum balance equations due to the flow of matter and gravitational radiation  through the screen.
We establish that the gravitational screen possesses analogs of {surface tension}, {internal energy} and {viscous stress tensor},
while the conservations are analogs of non-equilibrium balance equations for a viscous system.
This gives a precise correspondence between gravity in finite regions and non-equilibrium thermodynamics.
\end{abstract}
\maketitle

%
%\title{First order gravity: Immirzi parameter, Energy, boundary terms, first law and boundary symplectic structure}
%\author{Laurent Freidel}\email{lfreidel@perimeterinstitute.ca}
%\affiliation{Perimeter Institute for Theoretical Physics, 31 Caroline St, N, On N2L 2Y5,
%Waterloo, Canada}

%\rightline{VPI-IPNAS-09-04} 

%\renewcommand{\thefootnote}{\fnsymbol{footnote}} 
%\date{\today}
%\begin{abstract}
%We perform the covariant Hamiltonian analysis of gravity in the first order formalism
%including the contribution from  the Immirzi parameter.
%We construct the Hamiltonian and evaluate the boundary energy.
%
%\end{abstract}
%%\pacs{}
%
%\maketitle

%%%%%%%%%%%%%%%%%%%%%%%%%%%%%%%%%%%%%%%%%%%%%%%%%%%%%%%%%%%%%%%%%%%%%%%%%%%%%
%\section*{Outline of the paper}
%%%%%%%%%%%%%%%%%%%%%%%%%%%%%%%%%%%%%%%%%%%%%%%%%%%%%%%%%%%%%%%%%%%%%%%%%%%%%%%%%
\section{Introduction}

Unlike any other interactions, gravity is fundamentally holographic.
 This fundamental property of Einstein gravity manifests itself more clearly 
 when one tries to define a notion of energy for a gravitational system. 
It is well known that no local covariant notion of energy can be given in general relativity.
  The physical reason can be tracked to the equivalence principle. 
  Illustrated in a heuristic manner, 
  %this principle can be used to get rid of the gravitational field on a given world line of spacetime. Namely, 
  a free falling point-like particle does not feel any gravitational field, so no gravitational energy density can be identified at spacetime points.
 A more radical way to witness the holographic nature of gravity, comes from the fact that the Hamiltonian of general relativity coupled to any matter fields, exactly vanishes for any physical configuration of the fields. 
 If one asks what is the total energy of a closed gravitational system with no boundary, the answer is that it is zero for any physical configurations. This is a mathematical consequence of  diffeomorphism invariance.  It naively implies that the gravitational energy density vanish. 
 
 A proper way to accommodate this, is to recognize that a  
 notion of energy can only be given once we introduce a bounded region  of 
  space together with a time evolution for the boundary of this region.
   The time evolution of this boundary span a  timelike world tube equipped with a time foliation. 
   We will call such boundaries equipped with a timelike foliation, gravitational {\it screens}.
   They will be the subject of our study which focuses on what happen to a gravitational system in a finite bounded region.
 In the presence of gravity, the total energy of the region inside 
 the screen comes purely from a boundary screen contribution and the bulk contribution vanishes.
   In that sense,  energy cannot be localized but it  can be quasi-localized, i-e expressed as a local  surface integral  on the screen.
    
    The goal of this paper is two-fold. First, we revisit the definition and key properties of the energy and momenta associated with  regions bounded by  screens.
      We focus first on the canonical energy associated with the Einstein-Hilbert Lagrangian.
    The key point we want to stress, is that the screen  energy  density is given by a notion of surface acceleration.
    The other key point is that this energy is the sum of a translational energy and a boost energy.
    The boost energy is entirely due  to the presence of   boundary degrees of freedom 
    associated with the presence of the screen.
    The translational energy on the other hand is due to the usual gravitational degrees of freedom.
    It  is proportional to the screen radial acceleration and it is the gravitational analog of {\it Gibbs energy}. The boost energy density is on the other hand proportional to 
     the difference between  the acceleration of screen observers 
    with the Newtonian acceleration of Eulerian observers. 
    
    This allows to show that the gravitational screen possesses a {\it surface tension}  $\sigma$ proportional to its 
    inward radial acceleration \cite{deGennes}. It also shows that the screen  possesses a {\it Newtonian potential} $\phi$, whose gradient defines the acceleration of Newtonian (i-e Eulerian) observers.
    We also study the dependence of the energy on the change of boundary Lagrangian and show that the Legendre transform of the canonical energy defines  a notion of gravitational {\it internal energy}.    The density of internal energy  is found to be proportional to the radial expansion (i-e the screen's extrinsic curvature). In order to understand in a thermodynamical fashion all the elements entering the definition of energy and its variation we develop in the core of the paper a description of a  2+ 2 decomposition of the gravitational field.

    The second purpose of this paper is  to establish the law of time evolution of the screen energy, together with the computation of the anomaly appearing in the Poisson constraint algebra.
    This anomaly being due  to the presence of the screen. These constitute our main results.

    For a general screen the energy is not conserved in time.
    Gravitational and matter energy can flow in and out of the region bounded by the screen and 
    in the second part of the paper we focus on deriving the equation that governs this dissipative process.
    What is remarkable is the fact that the gravity equation of motion projects themselves onto the screen as the equation of non-equilibrium thermodynamic for a continuous isotropic and non elastic media, with  one component, i-e a general fluid. We will show  that, under this analogy, the screen possesses a {\it viscous stress tensor }$\tau$ proportional to the radial deformation tensor.
    
  This implies that the gravitational screen is exactly described as a thermodynamical system out of equilibrium. Lets  denote by $U$ the total  {\it internal energy} of the screen, defined as the integral of the radial expansion.
  The gravity equations holographically project themselves onto the screen as the first law of non-equilibrium thermodynamics which reads
  \be
  \rd U = \sigma \rd A + \delta E_{\mathrm{M}} +\delta E_{\mathrm{N} } + \delta Q
  \ee
  where $A$ is the area of the screen.
  The first term is a work term due to the presence of surface tension $\sigma$ and give the energy cost of expanding the screen area.
   $\delta E_{\mathrm{M}}$ is the work due to matter entering or leaving the region inside the screen, $\delta E_{\mathrm{N}}$ is the Newtonian work. The Newtonian energy comes from the coupling of the Newtonian potential $\phi$  to the screen's inertial mass which appears to be given by 
   the internal energy. So $U$ acts as the inertial mass for the non relativistic system described by the screen.
  At last $\delta Q$ represent the internal heat production of the screen due to viscous forces. 
  It is proportional to the contraction of the viscous stress tensor with the velocity gradient, i-e the time derivative of the screen's metric.
This term  represents the dissipation of energy due to the gravitational wave production.
It shows that gravitational wave energy is account for as heat dissipation.
We have seen already  that  the total energy of a closed isolated gravitational system is always zero.
From the thermodynamical point of view we can  understand this cancellation as a balance between work 
which represents usable energy and heat.
The  workable energy  comprises of matter energy, Coulombic energy and tension energy while the heat comprises of gravitational radiation. This equation is presented in section \ref{ebal}.
  
  Our formalism is valid for an {\it arbitrary} timelike screen. The usual situations studied in the literature 
  are often limiting situations obtained by specializing the screen to be either at infinity or along a black hole horizon.
  Black Hole  event horizons are surfaces that can be represented as a null limit of timelike screens. In this particular limit and under conditions of equilibrium, the first law presented here reduces to first laws established for black holes.
  However, the first law presented here is in many way more general. First it includes a Newtonian term and more importantly it shows that general screens possess an internal energy. The variation of this quantity is usually set up to vanish  in equilibrium situation, but it will not in a general non-equilibrium situation.
  
 Also, it shows clearly that the surface gravity do not appear naturally in the first law as a  temperature of the screen but as its {\it pressure} or negative {\it surface tension}. 
 Wether the identification of surface tension and temperature, valid for the very particular case of  a killing bifurcated horizon surface \cite{HT,UT}, extends or not for a more general screen, is an open question beyond the scope of this paper.
 In 4d gravity, the screen is 2-dimensional, therefore the usual pressure work term $-p_{3d}dV$ reads $-p_{2d} \rd A=\sigma \rd A$.
 This term is often misinterpreted for an entropy production term but it is not in this general context,
  it is the usual  work term due to change of the size of the system.
 
 The viscous entropy production term in non-equilibrium thermodynamics \cite{degroot} is due to the presence of a viscous stress tensor and related to the production of heat and internal dissipation.
 In our case we can clearly identify it with the production and transport of gravitational waves.
 So entropy production for the screen viewed as a thermodynamical system is the left-over signature of dynamical gravity.
 
 We also consider in this work the equation governing the conservation of the screen momenta.
 We establish that the momenta density is proportional to the so-called normal connection and we write explicitly the equation governing the non conservation of screen momenta.
 This equation confirms the thermodynamical interpretation given for the energy conservation.
 In particular it confirms that the screen possesses a surface tension proportional to its radial acceleration and an internal energy that acts as an inertial mass for the Newtonian potential.
 One finds that the screen acceleration is due to several term.
 Schematically,
 \be
 \delta P = \rd \sigma + \rd \dd \tau + F_{\mathrm{M}} +F_{\mathrm{N}}.
 \ee
 The first term is the ``Marangoni'' force \cite{Marangoni} due to the presence of surface tension gradients,
 the second one is the viscous force due to the presence of the viscous stress tensor $\tau$  (here proportional to the radial deformation tensor),  we also have a Newtonian force $F_{\mathrm{N}}= -U\rd \phi$ due to the Newtonian potential and finally a force $F_{\mathrm{M}}$ due to the transfer of momenta from the matter to the screen.

The goal of this paper is to give a self contained presentation of the construction of the symplectic potential,
gravitational energy, dependence on the boundary term, and the 2+2 formalism which is a key technique used here.
These subjects have been all touched on and developed on many instances in the literature, but in a scattered manner that we try to reunite here. We also want to give a unified presentation of the thermodynamical interpretation of the different form of energy that appears in the gravitational context.

The variational principle for gravity and construction of the symplectic potential has been developed in 
several instance by Regge, Teitelboim and more recently Wald, Iyer and Brown and York \cite{Var1, Var2, Var3}. The notion of quasi local energy has also been developed by many additional authors \cite{E1,E2,E3,E4,E5,E6,EReview}.
The two  definitions which  have attracted  most of the attention are the definition of Brown and York \cite{E1} which correspond as we will see to what we call the {\it internal energy} and the definition of Iyer and Wald \cite{Var2}
which corresponds  to our gravitational {\it Gibbs energy}. Both are canonical energies.
These definitions have been extended to non orthogonal boundaries in 
\cite{EE1, EE2, EE3, EE4}.
The presence of boundary degrees of freedom  has been introduced in the Lagrangian context as boundary terms needed to extend the Gibbons-Hawking prescription \cite{2dboundary, EE1}
and then appreciated first by Carlip and Teitelboim 
\cite{CT} in the Hamiltonian  context, as introducing a new canonical pair.
The notion of boost energy appears in recent  works related to quantum gravity \cite{Massar, perez,bianchi} even if its  relevance to the total energy has not been described precisely previously.
In our presentation we develop in great detail the 2+2 formulation of gravity, emphasizing the importance of the foliations scalars and detailing the accelerations. Some elements  but not all  of this decomposition appears in \cite{22,Hayward,Cao,Gourgoul}.

Our work is deeply inspired by the membrane paradigm as developed by Price, Thorne \cite{Price,Thorne} and Damour \cite{Damour} and can be viewed as a 
full extension of this program for a general timelike screen.
The main idea of this program and of our work being that one can replace the gravitational degrees of freedom 
inside a screen by boundary degrees of freedom on the screen.

Our work is also inspired by the beautiful developments associated with trapping,
isolated, dynamical and slowly evolving horizons \cite{trapping,IsolatedH,DynamicalH,slowlyH,Gourgoul2, Jaramillo}.
One key difference being that a dynamical horizon is a space like surface and cannot therefore be understood as a physical membrane. 
The first law and momenta conservation law that we write are nevertheless related to 
laws satisfied by these objects.

\section{Hamiltonian analysis and energy}
In this section we present the construction of the symplectic potential for gravity,
the presence of bulk and boundary degrees of freedom, the canonical Hamiltonian and its thermodynamical interpretations.
\subsection{ Boundary variations and conventions}

Our conventions are that the metric $g_{\a\b}$ possesses a signature $(-+++)$, its covariant derivative is denoted by $\N_{\alpha}$,  and the curvature tensor is defined to be
$ [\N_{\alpha},\N_{\beta}] v^{\mu} = R^{\mu}{}_{\nu \alpha \beta} v^{\nu}$.
In units where $ 8\pi G =1$, the gravity Lagrangian is given by
\be
L_{G}=\f12 \sqrt{|g|} g^{\a\b} R_{\a\b}.
\ee
The convention for matter fields are as follows: The scalar field matter Lagrangian is given by 
\be
L_{m} = -\sqrt{|g|}\left(\frac12 g^{\a\b} \partial_{\a}\phi \partial_{\b}\phi + V(\phi) \right),
\ee
and its energy momentum tensor by $ T_{\mu\nu} = - \frac{2}{\sqrt{|g|}} \frac{\d S_{m}}{\d g^{\mu\nu}}$,
where $S_{m}=\int L_{m}$.

In the following we will  use  notations,  often used in the relativity literature e.g \cite{3+1}, that allows to limit the number of indices contractions.
IN these notations a  vector with components $n^{\mu}$ is denoted by bold face letter $\n\equiv n^{\mu}\partial_{\mu}$,
the corresponding one form obtained by lowered the indices with the spacetime metric is denoted $\un{n}=n_{\mu}\rd x^{\mu}$.
Single contraction of vectors are denoted with a dot $ n^{\mu} t_{\mu}= \n\dd \t$ double contraction with a double dot
$ \sigma_{\mu\nu}\sigma^{\mu\nu}= \bm{\sigma}\!:\!\bm{\sigma}$.
The Lie derivative along a vector $\n$ is denoted $\cL_{\n}$ the interior product of a vector with a 
one form is denoted $\i_{\n}$, $\i_{\n}\un{t}= \n\cdot \t$.

Finally let us recall that the Gauss law is given by 
$$ \int_{M} \sqrt{|g|} \N_{\mu} v^{\mu} = \int_{\partial M } v^{\mu} \epsilon_{\mu}$$ where the surface element is given by $\epsilon_{\mu} = \sigma n_{\mu} \sqrt{h} \rd^{3}x$ for space like or timelike boundary. Here $n_{\mu}$ is the outgoing unit normal to the boundary, $\sigma \equiv n^\mu n_{\mu}$ is negative for a space like hyper surface and positive for a timelike hyper surface, and $ h_{\mu \nu} = g_{\mu\nu}- \sigma n_{\mu}n_{\nu}$ is the induced metric on the boundary. In the following we will also use the notation $\epsilon \equiv \sqrt{|g|} \rd^{4}x$ for the top form. More details on these is given in the appendix \ref{fid}.

\subsection{The setting: Observers, screens and foliation}
\begin{figure}[h!]
  \caption{Foliation and screens}
  \centering
    \includegraphics[width=0.8\textwidth]{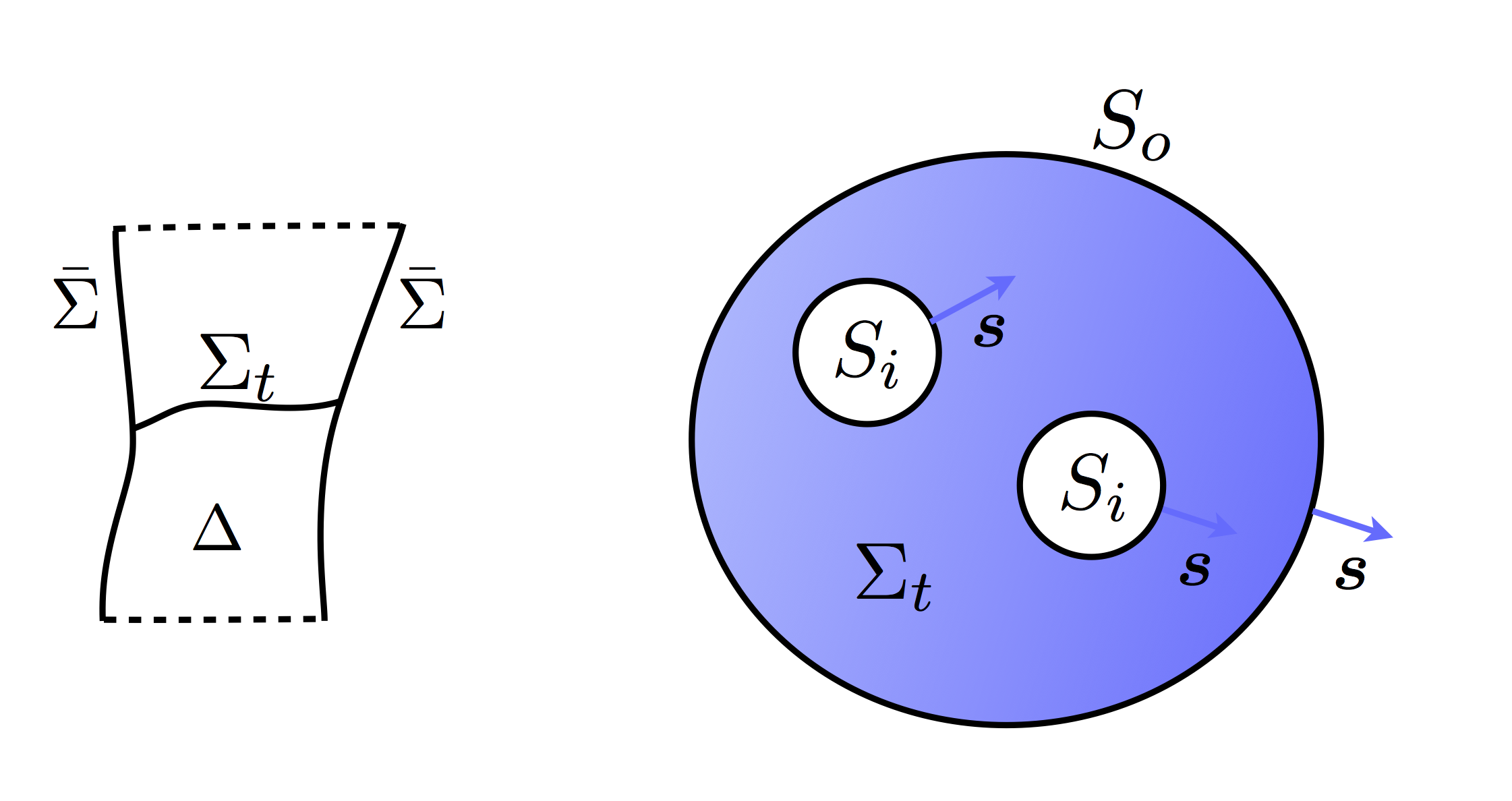}
    \label{folscreen}
\end{figure}

The setting we are interested in is the study of a connected region of space-time denoted $\Delta$ which possesses
a global foliation and which also possesses timelike boundaries called {\it screens}.
The screens are assumed to have the topology $S^{2} \times \mathbb{R}$. We also assume that there is 
one screen $\overline{\Sigma}_{o}$ which can be identified with the outer screen, while there a  (possibly empty) set of interior screens $\overline{\Sigma}_{i}$. 
This situation is pictured in figure \ref{folscreen}.
Our analysis is valid for a general set up of screens but it will be convenient at times to restrict to the case where there are only one outer and one interior screen. 

 The leaves of  the foliation are denoted by  $\Sigma_t$, they are the level set of a given spacetime time function $T(x)$,with value $t$.
%$\Sigma_t = \{ x\in M/ T(x)=t\}$.
The unit normal to $\Sigma_t$ is denoted by $\n$ and satisfies $\n\dd\n=-1$.
We denote by  $(h_{\mu\nu}, D_{\nu})$ the metric and covariant derivative on $\Sigma_{t}$.
They are related to the metric and connection
on the slices to the spacetime ones by the use of the orthonormal projector
\be
h_{\mu}{}^{\nu} = g_{\mu}{}^{\nu} + n_{\mu} n^{\nu}, \qquad h_{\mu}{}^{\alpha} \nabla_{\alpha} v_{\nu} =
D_{\mu}v_{\nu},
\ee
for a vector $\bm v$ tangent to $\Sigma_{t}$.
The time evolution is characterized by a time flow vector $\t=t^{\mu}\partial_{\mu}$ which can be decomposed in terms of a lapse and a shift 
\be
\t= N \n + \bm{M}.
\ee
This time flow vector is assumed to be parallel to the  boundary screens.
The characteristic property of this vector, that is $t^{\a}\partial_{\a}T=1$, implies  that the Lie derivative along $\t$ of any vector tangent to $\Sigma_{t}$ is still tangent to $\Sigma_{t}$. This means that 
\be
h_{\alpha}{}^{\mu}\cL_{\t} n_{\mu} =0.
\ee
An  Eulerien observer \cite{Smaryork}, is  a fiducial observer static with respect to the foliation, whose 4-velocity is given by $n^{\mu}$.
The previous identity implies  that the acceleration of an Eulerien observer, is a vector tangent to $\Sigma_{t}$ given by the space derivative of the lapse function:
\be
 \nabla_{\n} n_{\mu} = \f{D_{\mu} N}{N}. 
\ee

The foliation leaves $\Sigma_{t}$ intersect the screens along 2-spheres denoted 
$S_{t}=\bar{\Sigma} \cap \Sigma_{t}$. 
The bulk foliation induces a foliation of the screens.  We will call a screen with a specific time foliation a {\it gravitational observer}.
The space-time metric can be decomposed in term of the 2d metric $\bm q$ on $S_{t}$ as 
$$ g_{\a\b} = q_{\a\b} + s_{\a}s_{\b} - n_{\a} n_{\b}$$ where  $s_{\a}$ is a unit spacelike vector
 tangent to $\Sigma_{t}$ but normal to $S_{t}$. We will always chose this vector to be directed outwardly, from the inner regions to the outer region.
 The time flow vector is then decomposed as 
\be
\t = N \n + M \s + \bm{\varphi} 
\ee
where $\bm{\varphi}$ is a 2d lapse vector tangent to $S_{t}$.
It will be convenient for us to introduce the normal time flow $\bht \equiv  N \n + M \s$.
This normal time flow is tangent to the screen and orthogonal to $S_{t}$ and therefore proportional to a unit timelike vector tangent to the screen and denoted $\bbn$.
It will also be useful for us to introduce the normal vector $\bot \equiv N \s + M \n$.
This vector is proportional to  $\bbs$ the unit normal to the screen going outwardly.
\begin{figure}[h!]
  \caption{Screen and foliation normals}
  \centering
    \includegraphics[width=0.4\textwidth]{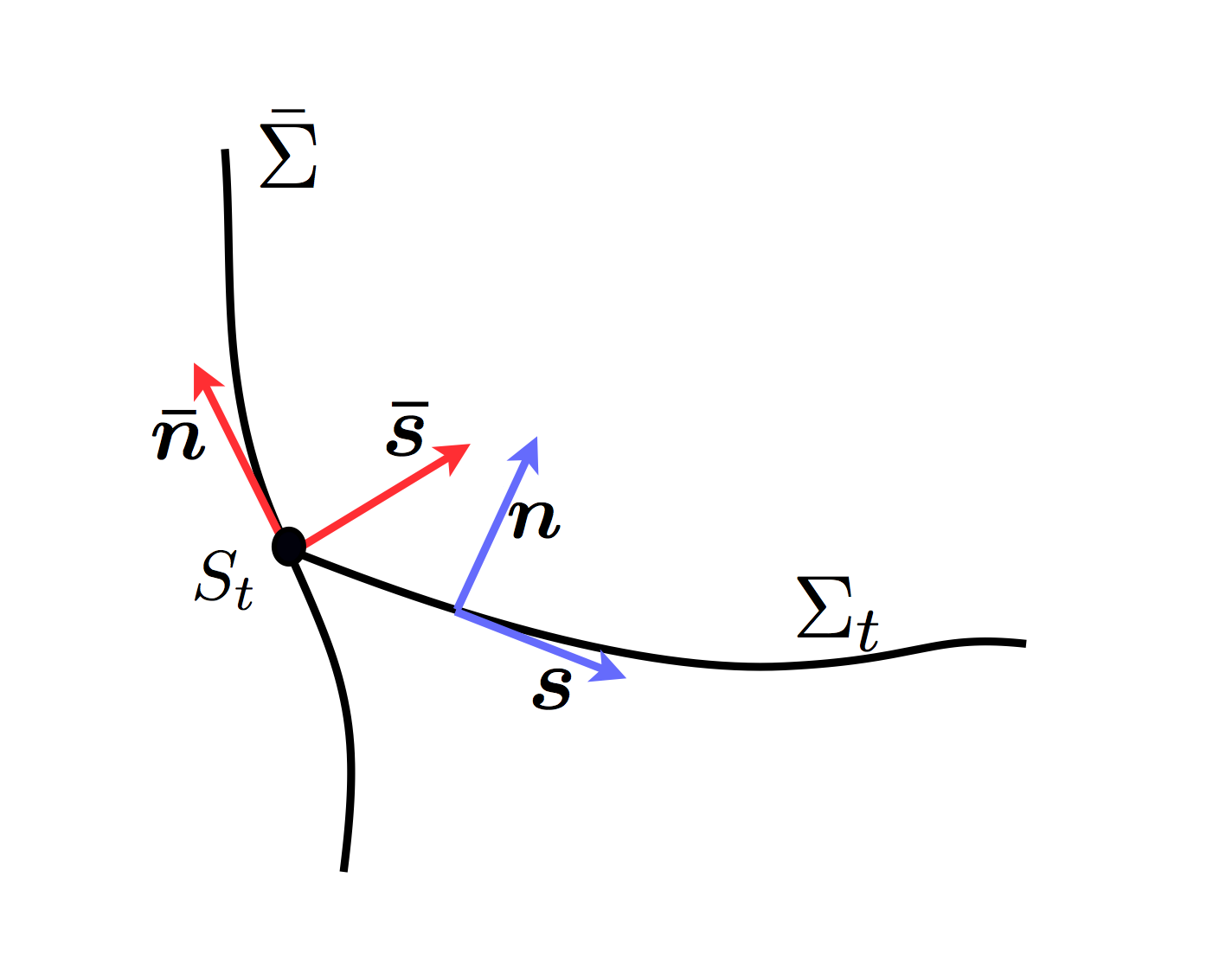}
\end{figure}
That is if we define  $N=\rho\cosh \beta$, $M= \rho \sinh \beta$, with
$\beta$ is the boost angle that relates the screen frame $(\bbn,\bbs)$ to the time foliation frame $(\n,\s)$.
We have 
$$
\bbn = \cosh \beta \n + \sinh\beta \s,\qquad \bbs = \cosh \beta \s + \sinh\beta \n.
$$
In the case where there is only  one outer and one interior screen, we assume that there are an additional foliation by timelike surfaces $\bar{\Sigma}_{r}$, which are the level surfaces $R(x)=r$ of a radial field,
which interpolates between the interior and outer screens.

%
%In order to get the symplectic structure of gravity we need to compute the boundary term in the 
%variation of the action.
%It is well known that the symplectic form is given by
%\be
%\bm{\Omega}=\delta \int_{\Sigma} \alpha, \mrm{with} \alpha =  \rd \delta L
%\ee
%$\alpha$ is the symplectic potential $\delta$ denotes variation on the space of fields, it is treated as a differential 
%in the sense that its square vanish\footnote{
%More formally suppose that we have a mapping $\alpha$ between a dimensional space $D$
%of deformation parameters and   the space of fields denoted ${\cal F}$. Given this map we can pull back  any functional of the field
%$F(\phi)$ onto functional on $D$. $\alpha^{*}F(d)= F(\phi_{d})$ where $\phi_{d}(x)$ denotes the fields
%labelled by the deformation parameter $d$.
%We can define $\delta $ to be the map $ \delta: {\cal F} \to \Omega^{1}({\cal F}) $ which is such that  its pull back on any deformation parameter space is the differential on $D$
%$\alpha^{*} \delta = \rd \alpha^{*}$.}$\delta^{2}=0$
%
%Note that it is in general possible to add to the action boundary terms.
%These boundary terms can be different for different boundary components since by definition the different components possess a different set of boundary fields.
%For instance in the gravity case some boundary components can be timmelike spacelike or null.
%Thus a more  general action in therefore
%given by
%\be 
%S =\int_{M} L + \sum_{i} \int_{\Sigma_{i}} \ell_{i}
%\ee

\subsection{Gravity symplectic potential}

Given a Lagrangian $L$ its symplectic potential 
is defined to be given by the boundary variation of the action.
We know that the bulk variation of the action defines the equation of motion and therefore the  on-shell
the variation of the Lagrangian is a pure derivative. It is given by 
\be
\delta S_{M}= \int_{M} \delta L  \,\hat{=} \int_{\partial M} \alpha,
\ee
where $\delta$ denotes variation on the space of fields and $\hat{=}$ denotes the on-shell evaluation.
The symplectic potential $\alpha$ can be itself decomposed into a bulk variation  $\alpha_{B}$ and a boundary variation $\alpha_{b}$.  
These boundary terms of the symplectic potential arises when
 the boundary of $M$ possesses corners, i-e co-dimension two manifolds  $S$ which separates 
two regions with different boundary conditions.
If we decompose the boundary of $M$ into co-dimension 1 components $\Sigma_{i}$ and corners $S_{ij}$
we can write the general on-shell variation of a Lagrangian on a manifold with corners as 
\be
\int_{M} \delta L \,\hat{=} \sum_{i}\int_{\Sigma_{i}} \alpha_{B} + \sum_{ij}\int_{S_{ij}} \alpha_{b}. 
\ee
The Lagrangian uniquely determines then the symplectic potential.
Once the symplectic potential has been identify, we can {\it uniquely} construct given a Lagrangian density $L$, the corresponding canonical Hamiltonian which is the canonical generator of time translation along $\t$. 
%In order to define it let us introduce an operation $I_{\t}$  similar to the interior product at the level of field variation. $I_{\t}$ is the operation of replacing the variation $\delta$ by the Lie derivative $\cL_{\t}$,
%that is $I_{\t} \delta \phi =\cL_{\t}\phi$, for any field. The canonical Hamiltonian 
It is given by
\be
H_{\t} \equiv  \int_{\Sigma_{t}}   ( I_{\t}{\alpha} - \imath_{\t}L )
%=-\f{1}{8\pi G} \int_{\Sigma_{t}}   \sqrt{h} ({\alpha}_{\t\n} + \frac{(\t\cdot \n)}2 R)
\ee
where $ I_{\t}\alpha \equiv I_{\t}{\alpha}^{\mu}\epsilon_{\mu}$ is the symplectic potential evaluated for time variations $I_{\t}\delta \phi =\cL_{\t}\phi.$ $\imath_{\t}$ denotes the interior product of the vector $\t$ with the Lagrangian form $ \i_{\t}\epsilon = t^{\mu}\epsilon_{\mu}$.

The goal is now to evaluate explicitly the canonical Hamiltonian for  gravity. 
We start by the computation of  the gravity symplectic potential using a fundamental identity for 
its evaluation.
This calculation appears in some form in many references, see e.g \cite{Var2, Var3, EE1} we present here for completeness and clarity
the calculation as this will set our notations and clarify what assumptions are made in its construction.
\subsection{A fundamental variational identity}

From the  definition of the Ricci tensor, and  using the expression of the variation of the connection 
$ \delta \Ga^{\mu}_{\alpha \beta} = \f12 g^{\mu\nu}(\N_{\alpha}\delta g_{\beta \nu} + \N_{\beta}\delta g_{\alpha \nu}-\N_{\nu}\delta g_{\alpha\beta})$,  we obtain that the Ricci tensor variation  is given by 
\be
\delta R_{\alpha\beta} = \N_{\mu}\delta \Ga_{\alpha \beta}^{\mu} - \N_{\alpha}\delta\Ga_{\mu\beta}^{\mu}.
\ee
Therefore we conclude that 
\be
\delta L =  
\f12 \sqrt{|g|} \,G_{\alpha \beta}  \delta g^{\alpha\beta}
+
\sqrt{|g|}\, \N_{\mu} \alpha^{\mu},
\ee
where $G_{\alpha\beta}$ denote the Einstein tensor and the symplectic potential vector is
%\footnote{If we introduce the variation $\Delta^{\mu}_{\nu}\equiv g^{\mu\a} \d g_{\a\nu}$, we can 
%write the symplectic form in a condensed form as 
%$\alpha^{\mu} = \N^{[\nu} \Delta^{\mu]}_{\nu}$, where the bracket denotes antisymmetrisation.}
\be\label{thetad}
\alpha^{\mu} =\f12\left( \d \Ga_{\a\beta}^{\mu} g^{\a\beta} - \d\Ga_{  \a \beta}^{\beta} g^{\mu\a}\right) = \nabla_{\nu} \alpha^{\mu\nu},\mrm{with} \alpha^{\mu\nu} \equiv \f12 \left(g^{\mu \alpha}g^{\nu\beta} -g^{\mu\nu} g^{\alpha\beta}\right)\delta g_{\alpha\beta},
\ee
while the symplectic potential is $ {\alpha} = \alpha^{\mu}\epsilon_{\mu} $.

In order to give an explicit expression for the symplectic potential, for a slice normal to the one form $n_{\a}$, let us introduce the induced metric tensor $ h_{\a\b }\equiv g_{\a\b} - \sigma n_{\a}n_{\b}$ , where $\sigma = n_{\a}n^{\a}$ is the signature. It is negative for a spacelike slice and positive for a timelike one\footnote{ We are initially interested in the case of a timelike slice but our formalism also works for a space like one. This will be needed in a later section.}. 
We also introduce the extrinsic curvature tensor 
$K_{\n \a\b} \equiv \frac12 \cL_{\n} h_{\a \b} = h_{\a}{}^{\a'}h_{\b}{}^{\b'} (\N_{\a'}n_{\b'})$,
and denotes its trace by $K_{\n}\equiv K_{\n \a\b}h^{\a\b}$.
We finally introduce the acceleration vector $ a_{\n}^{\mu} \equiv -\sigma \N_{\n} n^{\mu}$.
From the definition, it is easy to check  that $ \N_{\a}n_{\b} = K_{\n \a\b} - n_{\a} a_{\n \b}$.
From now on we also denote $\alpha_{\n} \equiv \alpha^{\mu} n_{\mu}$ so the 
symplectic potential reads ${\alpha} = \sigma \sqrt{h} \alpha_{\n}$.

 We now 
 establish, using the definition (\ref{thetad}), that
\be\label{divid}
\d(\N_{\a}n^{\a} +\N^{\a}n_{\a}) 
= \N_{\a}\d n^{\a} +\N^{\a}\d n_{\a} + \d g^{\a\b} \N_{\a}n_{\b} -  2 \alpha_{\n}.
\ee
Using the identity $ \N_{\mu}(h^{\mu}{}_{\nu} \d n^{\nu})= D_{\mu}(h^{\mu}{}_{\nu} \d n^{\nu}) + a_{\n\mu}\d n^{\mu},$ where $D_{\mu}$ is the derivative on the slice compatible with the induced metric $h$, we can expand the first two terms of the RHS as 
$$
\N_{\a}\d n^{\a} +\N^{\a}\d n_{\a} = D_{a} \d_{\n}^{a} + a_{\n a} \d n^{a} + a_{\n}^{a}\d n_{a}.
$$
Where we have introduced 
${\delta}_{\n}^{a} \equiv (h^{a}{}_{\nu} \d n^{\nu}+ h^{a \nu} \d n_{\nu}).$
%We first expand as follows
%$$\d g_{\a\b}n^{\b}=\d n_{\a} - g_{\a\b}\d n^{\b},\qquad 
%\N_{\mu}(h^{\mu}{}_{\nu} \d n^{\nu})= D_{\mu}(h^{\mu}{}_{\nu} \d n^{\nu}) + a_{\mu}\d n^{\mu},$$
Finally, using the definition of $K_{\n}$  and the variational identity $\d g_{\a\b}n^{\b}=\d n_{\a} - g_{\a\b}\d n^{\b}$
we establish that 
\be
\delta g^{\a\b} \N_{\a}n_{\b}= \d h^{ab} K_{\n ab} + a_{\n}^{a}\d n_{a} - a_{\n a} \d n^{a}.
\ee   This allow us to  establish after rearrangements,  the fundamental identity:
\bea\label{theta1}
\boxed{ -\sqrt{h} \,\alpha_{\n}=  \d \left( \sqrt{h} K_{\n}\right) +  \frac{\sqrt{h}}{2}(K^{ab}_{\n}-h^{ab} K_{\n}) \delta h_{ab} - \sqrt{h} a^{a}_{\n}\d n_{a}-
 \frac{\sqrt{h}}2D_{a}{\delta}_{\n}^{a}
 %\left(h^{\mu}{}_{\nu} \d n^{\nu}+ h^{\mu \nu} \d n_{\nu}\right)
 .}
\eea
where ${\delta}_{\n}^{a} \equiv (h^{a}{}_{\nu} \d n^{\nu}+ h^{a \nu} \d n_{\nu})$.
This is the expression we were looking for.
\subsection{Bulk and boundary contributions}

The first term in \Ref{theta1}    is a total variation,
 therefore it  does not contribute to the symplectic structure
 even if it does affect the symplectic potential.
 This term is the variation of the celebrated Gibbons-Hawking boundary term \cite{Gibbons-H}.
 
The second term and third terms  determine the bulk symplectic structure.
First it shows the well known fact that 
$$ \boxed{\Pi^{\mu\nu}\equiv\sqrt{h}(K^{\mu\nu}_{\n}-K_{\n} h^{\mu\nu})/2}$$
 is the momentum conjugated to $h_{\mu\nu}$.
Since $ \Pi^{\mu\nu} h_{\mu\nu} = -\sqrt{h} K_{\n}$
we have that
\be
\delta (\sqrt{h} K_{\n}) +\Pi^{\mu\nu}\delta h_{\mu\nu}
= - \delta \Pi^{\mu\nu}  h_{\mu\nu}.
\ee
The third  term given by $- \sqrt{h} a^{\mu}_{\n}\delta n_{\mu}$ depends on the parameters labelling the foliation.
Since $n_{\mu}=-N \p_{\mu}T$ and $a_{\n \mu}= D_{\mu}N/N $,  this component of the symplectic potential can be written as 
$$
 -\sqrt{h} a^{\mu}_{\n}\delta n_{\mu}= \sqrt{h}  D^{\a}N   D_{\a }\delta T
 = - \sqrt{h} (D_{\a}D^{\a}N)  \delta T + \sqrt{h} D_{\a}( D^{\a}N \d T) .
$$
This shows that the momentum conjugate to the foliation time $T$ is given by
\be
\boxed{\Pi_{T} = - \sqrt{h} (D_{\a}D^{\a}N).}
\ee
The lapse depends on $T$ and the metric via $N(T) = \left(- \partial_{\mu}T g^{\mu\nu}\partial_{\nu}T\right)^{-\f{1}{2}}$.
We can ignore this contribution to the symplectic potential when we consider  hypersurface orthogonal deformations 
that modify the fields without changing the foliation, that is deformation such that $h_{\alpha}{}^{\mu}\delta n_{\mu}=0$.
On the other hand,
for a arbitrary  spacetime diffeomorphism $\bm \xi$,
  %does change the value of $\n$  since 
  $h_{\alpha}{}^{\mu}\cL_{\bm \xi}n_{\mu}$ doesn't necessarily vanish.
This means that not all diffeomorphisms can be represented as a hypersurface orthogonal deformations 
and therefore can be implemented canonically in terms of a symplectic structure that depends only on $(\bm{h},\bm{K})$.
The condition $h_{\alpha}{}^{\mu}\cL_{\bm\xi}n_{\mu}=0$ is equivalent to $\xi^{\mu}n_{\mu}=c(T)N$ where $c$ is a function that depends only on $T$.
Indeed
\be
h_{\alpha}{}^{\mu}\cL_{\bm{\xi}} n_{\mu} = D_{\alpha}(\bm{\xi}\dd \n) - a_{\alpha}(\bm{\xi}\dd \n) = ND_{\a}c.
\ee

In summary, for a general variation,  the bulk symplectic potential is  given by
\be\boxed{
 \Theta_{\Sigma} = \int_{\Sigma_{t}} \alpha^{\mu}\epsilon_{\mu} =
 -\int_{\Sigma}   h_{\mu\nu}\delta \Pi^{\mu\nu}  + \int_{\Sigma} \Pi_{T} \d T.}
 \ee
The second term vanish for surface orthogonal variations.
In this case  the Bulk canonical variables are therefore the usual pairs $(\Pi^{\mu\nu},g_{\mu\nu})$,
with $\Pi^{\mu\nu}=\sqrt{h}(K^{\mu\nu}-h^{\mu\nu}K)$, if one restrict to foliation preserving variations.
They and also include $(\Pi_{T}, T)$ for a general variation, with
$\Pi_{T}= -\sqrt{h} \Delta N.$

\subsection{Boundary symplectic potential}

What appears from this computation is that we also have boundary degrees of freedom that contributes
to the symplectic potential. These arises since we are considering finite boundaries.
We restrict to variations that do not change $T$ on the screen. In this case the boundary symplectic potential takes the form
\be
\Theta_{S} =-\frac12  \int_{S}  \sqrt{q}  \bm\delta,\quad \mathrm{with} \quad \bm\delta \equiv ( s^{\mu}\d n_{\mu} + s_{\mu}\d n^{\mu})\label{boundsymp}
\ee
In order to understand the nature of this term, let us introduce a time coordinates $T$ that label the slices $\Sigma$ and a  radial coordinate $R$ that label the position of the screen.
Since $n_{a}$ is a  one-form normal to the slice it is proportional to $\rd T$, its normal radial unit vector 
$s_{a}$ will be proportional to $\rd R$ only if the slicing is orthogonal to the screen. But in general it is given by a linear combination of $\rd T $ and $\rd R$. We therefore need 3 foliations scalars to characterize the position of the screen and slicing, these are given by a time lapse $\rho$, a space lapse $\tau$ and a boost angle $\beta$, they are defined by:
\be
\underline\n  = -\rho\cosh \beta \rd T, \qquad \underline\s  = \tau \rd R + \rho \sinh\beta \rd T.
\ee
where $\underline\n \equiv n_{a}\rd x^{a}$.
The boost angle $\beta$ is the angle needed in order to rotate the slicing frame into the screen frame, since  $\underline\bbs\equiv \cosh\beta \underline\s + \sinh \beta \underline\n \propto \rd R $. The meaning of $\rho,\tau$ comes from the fact that
$\rho \rd T$ measure the proper time as it flows on the screen while $\tau \rd R$ measure the proper radial distance  on the slices $T=cste$.
The screen velocity is  the velocity of the screen as seen by an Eulerian (static) observer. 
Such an observer is characterized by the fact that it doesn't move on the slicing hence $ s_{a}\dot{x}^{a} =0 $ which implies that it posses a radial velocity $- v_{R}$ where
\be
v_{R} = \frac{\rho}{\tau}\sinh\beta.
\ee $v_{R}$ is the velocity of the screen relative to static observers.
The explicitly calculation of $\bm\delta$ is given in the appendix and also in  section \ref{22}. The result is that 
\be
\bm{\delta} =-\left( \d \beta + \tanh\beta\left(\frac{\delta\rho}{\rho}-\frac{\delta \tau}{\tau} \right)\right)
= - \tanh\beta \frac{\d v_{R}}{v_{R}}.\label{dvar}
\ee
So we see that for particular variations where  the rate of change of the time lapse is equal to the rate of change of the space lapse i-e $\delta\rho/\rho =\delta\tau/\tau$, this is simply equal to the variation of the boost angle. In this case the boundary symplectic structure  is simply
\be
\Theta_{S} = \f12 \int_{S}  \sqrt{q}  \delta \beta.
\ee
This shows that the surface area element $\sqrt{q}$ and the boost angle $\beta$ are in this context conjugated variables. This was first recognised by Carlip and Teitelboim \cite{CT}.
This statement is however not generally true as we just saw since $\d \bm \d \neq 0$ in general. For a general variation we get instead  that the symplectic structure is of the form
\be
\Theta_{S} = \f12 \int_{S}  \left(\frac{\tau \sqrt{q}}{\rho \cosh \beta}\right) \delta v_{R}.
\ee 
so that $v_{R}$ and the rescaled area element $ \frac{\tau \sqrt{q}}{\rho \cosh \beta}$ are the boundary conjugate variables.

Even if $\bm \d$ cannot be written as the variation of a boost angle, it is still of interest to introduce the notion of boost angle associated with  one particular variation. 
We naturally choose $I_{\t}\delta =\cL_{\t}$ so that $\eta$ is defined by
\be\label{etadef}
 \cL_{\t}\eta \equiv  ( s^{\mu}\cL_{\t}n_{\mu} + s_{\mu}\cL_{\t}n^{\mu}) = I_{\t}\bm{\delta}.
\ee
As we will see this angle naturally enters the definition of the total energy.

\subsection{Canonical gravitational Energy}\label{Cenergy}
The gravitational Hamiltonian which is the canonical generator of time translation is given by
\be
H_{\t}^{G} \equiv \f{1}{8\pi G} \int_{\Sigma_{t}}   ( I_{\t}{\alpha}^{\mu}\epsilon_{\mu} - \imath_{\t}L )=
-\f{1}{8\pi G} \int_{\Sigma_{t}}   \sqrt{h} \left(I_{\t}{\alpha}_{\n} - \frac{(\t\cdot \n)}2 R\right),
\ee
where $ I_{\t}{\alpha}{}^{\mu}$ is the symplectic potential vector evaluated for variations $I_{\t}\delta \phi =\cL_{\t}\phi$.
Note that 
%$\alpha_{\t \n} \equiv \alpha_{\t}{}^{\mu} n_{\mu}$,and
 $N = -\t\dd\n$ is the lapse function given by $\sqrt{g} = N \sqrt{h}$.
From the previous section we know that 
\be
I_{\t}{\alpha}_{\n} =  \frac{1}{\sqrt{h}}h_{\alpha\beta}\cL_{\t}\Pi^{\alpha\beta} +\f12  D_{\a} \cL_{\t} n^{\a},  
\ee
where we have 
used that $h^{\mu}_{\alpha} \cL_{\t}n_{\alpha} =0$, since the time flow $\t$ preserves the foliation.
  Thus the bulk  term depending on the foliation do not enter the 
definition of $H_{\t}$.

From the definition of $\Pi$ and the Ricci-Codazzi identity \Ref{Rictr} we have that 
\bea
%-h_{\alpha\beta}\cL_{\t}\Pi^{\alpha\beta} 
I_{\t}{\alpha}_{\n}&=&\frac1{\sqrt{h}}\left(\cL_{\t}( h_{ab} \Pi^{ab} )- (\cL_{\t}h_{ab}) \Pi^{ab} \right)+ \f12 D_{a} \cL_{\t} n^{a}\nn\\
&=&   -\frac1{\sqrt{h}} \cL_{\t}(\sqrt{h}K) - \f12 (D_{a}t_{b} + D_{b} t_{a})(K^{ab}-h^{ab}K) + \f12  D_{a} \cL_{\t} n^{a}\nn\\
&=& -\left(\cL_{\t}(K)+  K^{ab}  D_{a}t_{b}  \right) + \f12  D_{a} \cL_{\t} n^{a}\nn\\
&=&  R_{\t\n} -  D_{a} \left( \N_{\t}n^{a}-\f12 \cL_{\t} n^{a}\right).\label{Conseq}
\eea
Integrating by part and using that 
\be
\int_{\Sigma}\sqrt{h} D_{a} v^{a} =\int_{S_{o}}\sqrt{q}(\bm v\cdot \s) - \sum_{i} \int_{S_{i}}\sqrt{q}(\bm v\cdot \s)\equiv \int_{\partial \Sigma}\sqrt{q}(\bm v\cdot \s)
\ee
where $S_{o}$ denotes the outer sphere boundary and $S_{i}$ the inner spheres, while 
$\s$ is a space like  directed towards the outer boundary.
This shows, that the canonical gravity Hamiltonian is given
%\footnote{Note that we use that 
%$\int_{\Sigma}\sqrt{h} D_{\alpha} v^{\alpha}=\int_{S_{o}}(\bm v\cdot \s)\sqrt{q} - \sum_{i} \int_{S_{i}}(\bm v\cdot \s)\sqrt{q} $ for $\partial \Sigma = S_{o}\cup_{i} S_{i}$, 
%for $\bm v$ space like and $\s$ directed towards the outer boundary as we choose.
%As a short hand we denote this $\int_{\partial \Sigma} \sqrt{q} (\bm v\cdot \s) $}
 by
\be
\boxed{H_{\t}^{G} \equiv   \f{1}{8\pi G} \int_{\partial \Sigma_{t}} \sqrt{q}  \kappa_{\t}   - \f{1}{8\pi G} \int_{\Sigma_{t}} \sqrt{h} G_{\t\n} }
\ee
We see that this energy contains a surface contribution and the value of the 
surface energy density is $\kappa_{\t}/8\pi G $ 
where $\kappa_{\t}$ is defined to be  
\be
\kappa_{\t} \equiv  \s\cdot \N_{\t}\n - \frac12 s_{a}\cL_{\t} n^{a}.
%\mrm{with}
%\d_{\t} \equiv \frac12 (s_{\a}\cL_{\t} n^{\a}- n_{a}\cL_{\t} s^{a}) .
\ee
Note that due to hyper surface orthogonality of the flow generated bt $\t$ we have that $ n^{a}\cL_{\t}s_{a}=0$ and using the definition of the dihedral angle \Ref{etadef}, we can  write the surface energy density as 
\be\label{kappa}
\kappa_{\t} \equiv  \s\cdot \N_{\t}\n - \frac12 \delta_{\t} \mrm{with}
\d_{\t} \equiv  (s_{a}\cL_{\t} n^{a}- n_{a}\cL_{\t} s^{a})=\cL_{\t}\eta .
\ee
As we will see in more detail later, the first term is the radial acceleration of the screen while the second is a boost acceleration.
The canonical energy of matter is given by\footnote{The matter momentum vector associated with a slice is given by
\be
p_{\mu}=\int_{\Sigma} T_{\mu}{}^{\nu} \epsilon_{\nu}= -\int_{\Sigma} \sqrt{h} T_{\mu \n},
\ee
while the energy of a slice is given by $ e= -\bm p\cdot \t$, the minus signs due to the choice of signature cancel.}  
\be
H^{M}_{\t} =  \int_{\Sigma_{t}} \sqrt{h} T_{\t\n} =  \f{1}{8\pi G} \int_{\Sigma_{t}} \sqrt{h} G_{\t\n}.
\ee
$ T_{\t\n}$ represent  the matter energy density $\Sigma$ as measured by an observer following the world line $\dot{x}^{\mu}=t^{\mu}$.
This shows that the total energy is simply given by a boundary term:
\be\label{E}
H_{\t} = H^{M}_{\t} + H^{G}_{\t} = \f{1}{8\pi G}\int_{\partial \Sigma_{t}} {\sqrt{q}} {\kappa_{\t}}.
\ee 
If we decompose $\partial\Sigma_{t} = S_{t o} \cup_{i} S_{t i}$ in term of its outer boundary and inner boundaries,we can express this energy can be written in terms of contribution for each boundaries
\be\label{cane}
 H_{\t} =\f{1}{8\pi G}\int_{S_{\t i}}^{S_{\t o}} \sqrt{h} \kappa_{\t}.
\ee
The outer boundary contributes positively and the inner ones negatively.
 $\kappa_{\t}$ is defined in \Ref{kappa} with $\s$ being pointing in the outer direction.
 
Let us emphasize here that this energy formula, presents two key features. First, it is quasi-local: it is non vanishing only on the boundary of the region of observation. This is a consequence of diffeomorphism invariance which implies that the bulk Hamiltonian vanish. In this sense gravity is naturally holographic.

Second, the energy depends on the choice of observer, that is not only the choice of screens, but also the choice of foliation of the screens. This second feature is not that unusual, for instance in special relativity different boosted observers possess different energies, however it is a feature that has led to a lot of resistance
since energy is usually associated with the study of stability and the usual point of view is that stability should be a property of a system, not a system and an observer.
There has always been attempts to define a unique notion of energy, The ADM energy is one example or the energy associated with black hole horizons. In each case it always amounts to choose a special type of observer, infinity with the flat slicing or the killing observer or the observer attached to a bifurcated surface.
These are beautiful and in the case of ADM lead to a result of positivity but they do not apply for a general space-time and 
if we want observer independence  we are left in a unsettling situation where special observers can be find only in special situations, while in others no notion of energy is available. Our point of view is that
in the general case we have to give  up such attempts and embrace the fact that the notion of energy and momenta is observer dependent. In this case the canonical energy is uniquely defined and given by \Ref{cane}.

\subsection{Decomposition of canonical energy}\label{dec}

What is remarkable about the  formula for the total energy is that is can naturally have a thermodynamical interpretation.
%It is customary to interpret  the surface energy density $\kappa_{\t}/2 \pi $ as a local temperature 
%and the area element $\rd A/ 4G $ as a local entropy, we obtain an expression for the canonical energy as a entropy like  term given by
%\be
%H_{\t} =\int_{S} \frac{\kappa_{\t}}{2\pi}  \,\, \frac{ \rd A}{4G}.
%\ee
In order to see this, it is convenient to introduce a decomposition of the time flow vector into a boundary ``normal-time'' vector $\bht$, tangent to the screen $\bS$,  but normal to $S$ and a ``rotation'' vector $\bm{\varphi}$ which is tangential  to $S$.
That is we define 
\be
\bht = -(\t\cdot \n) \n + (\t \cdot \s) \s,\quad \mrm{and} \quad \t =\bht +\bm{\varphi}.
\ee
This decomposition is mirrored in the decomposition of the energy density $\kappa_{\t}=\kappa_{\bht}+\kappa_{\bphi}$.
Using this decomposition we can introduce the quasi-local mass and angular momenta\footnote{Strictly speaking it is truly an angular moment when the orbits of  $\bphi$ are closed circles of length $2\pi$ foliating the two sphere.
In general it should be understood as a momenta as we will see.}
$S$, they are given by
\be
M_{\bht} \equiv  2 \int_{S} \sqrt{q}\frac{\kappa_{\bht}}{8\pi G},\qquad 
J_{\bm\varphi} \equiv -  \int_{S} \sqrt{q} \frac{\kappa_{\bphi}}{8\pi G}, 
\ee
As we will see in the next section, the normalisation is chosen in order to reproduce, Komar mass and angular momenta formula
\cite{Komar} for space-times where $\t$ is a Killing field.
It also reproduces the Newtonian expression for the Newtonian mass as a Gauss law.
Finally, the formula for the mass also reproduce ADM mass formula if the screen is located at timelike infinity. It also reproduces Bondi energy \cite{Bondi} for a screen that asymptote null infinity.
We therefore see that for different choice of screen and time the canonical mass formula reproduces, Komar, ADM or Bondi.

One of the key feature of having  finite boundary is the appearance of boundary degrees of freedom\footnote{Let us emphasize that tho is a special feature of gravity that possesses a boundary symplectic structure.
This will {\it not} be the case for the matter fields theory with the notable exception of the theta term in Yang-mills theory.}
associated with the boundary symplectic potential $-\frac12 \int_{S} \sqrt{q} \delta_{\t}$. 
These degrees of freedom are new degrees of freedom $(\sqrt{q},\eta)$ not part of the usual gravitational degree of freedom $(h_{ab},K_{ab})$, they are entirely due to the presence of the boundary.
This dependence of the energy on the boundary degree of freedom shows up in the decomposition
of the surface energy density as a sum of two terms:
$
\kappa_{\t} = \gamma_{\t} -\frac12\delta_{\t},
$
where 
\be
\gamma_{\t}\equiv \s \N_{\t} \n,\qquad \delta_{\t} =\left( s_{\mu}\cL_{\t}n^{\mu}+s^{\mu}\cL_{\t}n_{\mu}\right).
\ee
This expression is written in terms of the foliation frame $(\n,\s)$. Since the energy is associated to the screen and the time evolution is parallel to the screen it is natural to look at this decomposition in the 
screen frame $(\bbn,\bbs)=\cosh\beta (\n,\s)+\sinh\beta(\s,\n)$.
where $\bbn$ is such that  $\bht =\rho \bbn$, $\rho$ being a screen lapse.
It is direct to check that $\kappa_{\t}$ is invariant under such change of frame and we can therefore write it as $
\kappa_{\t} = \bar{\gamma}_{\t} - \frac12 \bar{\delta}_{\t},
$
where 
\be
\bar\gamma_{\t}\equiv \bbs \N_{\t} \bbn,\qquad \bar\delta_{\t} =\left( \bs_{\mu}\cL_{\t}\bn^{\mu}+\bs^{\mu}\cL_{\t}\bn_{\mu}\right).
\ee
Here $\bar\gamma_{\t}= \gamma_{\t} +\cL_{\t}\beta$ is the radial acceleration of the screen,
while $\bar\delta_{\t}$ is a boost acceleration measuring the relative radial  acceleration of screen observers with respect to fiducial static observers.

Accordingly, the total energy $H_{\t}$ decomposes in a screen contribution $G_{\t}$ and a boundary contribution $h_{\t}$. These are given by
\be
G_{\t}\equiv \int_{S}\sqrt{q} \frac{\bar\gamma_{\t}}{8\pi G},\qquad 
h_{\t}\equiv - \int_{S}\sqrt{q} \frac{\bar\d_{\t}}{16\pi G}.
\ee
As we are about to see the screen energy $G_{\t}$, is the gravitational analog of the Gibbs energy\footnote{ The justification for this denomination is given in the next section comes from the fact that its variation involves only variation of the connection which are ``intensive'' variables.}.
The decomposition  of $\t =\bht +\bphi$ implies a natural decomposition of  Gibbs energy in terms of 
  the screen {\it surface tension} $\sigma_{\t}$  and the screen {\it momenta density}  $p_{\bm \varphi}$.
Explicitly $G_{\t}= - \int_{S} \sqrt{q}(\sigma_{\t}+p_{\bphi})$, with
\be
\boxed{
\sigma_{\t} \equiv - \frac{\bbs \N_{\bht}\bbn}{8\pi G},\qquad p_{{\bphi}}\equiv -\frac{\bbs\N_{\bphi}\bbn}{8\pi G}.
}\label{sp}
\ee
The understanding that $\sigma_{\t}$ is really the surface tension of the gravitational screen 
will be revealed when we establish the general first law.
The surface tension can be positive, generally for inner screens, or negative, generally for outer screens.
When negative it is better understood as a 2dimensional pressure for the screen, as in usual fluid systems where negative tension is pressure.

\begin{figure}[h!]
  \caption{Translation and boost slices}
  \centering
    \includegraphics[width=0.2\textwidth]{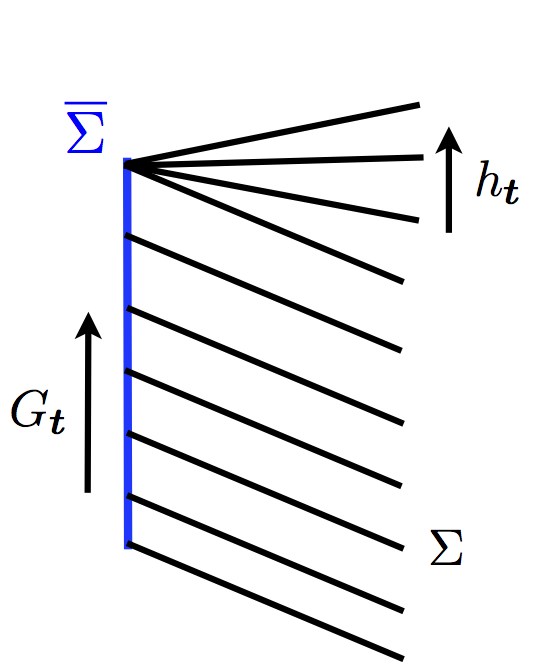}
    \label{trans+boost}
\end{figure}

The fact that the total energy is the sum of two different types of energy is unsettling at first.
There is however a beautiful geometrical interpretation of this fact.
The Gibbs energy $G_{\t}$ appears to be the canonical generator of translation along the screen, it corresponds therefore to the usual translational energy.
The second contribution $h_{\t} =\int_{S}\sqrt{q} \cL_{\t}\bar{\eta} $ appears to be the canonical generator of boost transformation at the screen. It doesn't generates translation and correspond to a boost energy.
This can be understood by the fact that a general motion of a foliation along a screen is characterized by 
translation and boost. We illustrate this in fig \ref{trans+boost}.
It is usual to associate a notion of energy for the generator of translation. It is less common however  to think of the generator of boost as an energy.
There is one context where this appears naturally, this is in the context where one computes entanglement energy \cite{Haag,entanglementH,MyersCasini}.
In this case given a space region $R$ with boundary $S$ and a vaccua state $|0\ket$ 
we associate to this data a density  matrix $\rho_{R} \equiv \tr_{\bar{R}}|0\ket \bra 0|$ where the trace is over all states that 
have support on the exterior of $R$. This density matrix can be written in terms of an operator  $\rho_{R}= exp(- 2\pi K_{R})$ where $K_{R}$ is the entanglement energy associated with the region $R$. This energy appears to be given by the  boost energy as is exemplified for instance in the context of the Unruh effect.

\subsection{Thermodynamical interpretation}\label{thermo inter}

Let us now discuss the thermodynamical interpretation of the canonical energy and its relation to mass and angular momenta
Since the 2-d metric $q_{AB}$ is conformally equivalent to the round sphere metric  
we denote by $\bm{\hat{\varphi}}$ a conformal killing vector whose close orbits have length $2 \pi$,
and we chose $\bm{\varphi}$ to be equal to $ \Omega \bm{\hat{\varphi}} $, where the angular velocity 
$\Omega$ is chose to be constant on $S^{2}$.
$J_{\bm{\hat{\varphi}}} \equiv J$ is the angular momenta of the screen and 
therefore the canonical energy is given by
\be
H_{\t} = \frac12 M_{\bht} - \Omega J_{\bm{\hat{\varphi}}}.
\ee
This decomposition is puzzling at first because of the factor $1/2$ and the minus sign.
The factor $1/2$ seems anomalous and there has been attempts in the literature to  fix this by 
adding a term that depends on a background structure \cite{Katz}. Let us now explain that this factor is instead welcome.
%This is however nothing but the expression of a generalisation of Komar mass formula\cite{}.

Lets suppose for a moment that both $\kappa_{\t}$ and $\Omega $ are constant on the horizon.
%and that $\bm \phi = \Omega \bm r$ where $\bm r$ has a closed orbit normalized to be $2\pi$.
%the vector flow is a Killing vector of a rotating black hole space time, 
%and that the screen is at the horizon, that is
%$\t = \bht + \Omega \bm{r}$ where $\bht$ is a time translation normalsie to be 1 at infinity and $\bm r$ is a rotational killing vetoer whose orbit as length $2\pi$.
In this case    the previous relation can be written as 
\be\label{GD}
\frac12 
M_{\bht} = T_{\t} S  + 
 \Omega {J}
\ee
where $S\equiv A/4G$, $J=J_{\bm \hat{\varphi}}$ and $T_{\t}\equiv {\kappa_{\t} }/{2 \pi }$.
Here and in order to ease the reader, I use the standard notations, valid in the Black hole case, which refer to the surface density energy as temperature and area as 
entropy even if it is not valid in the general context.
This is now naturally interpreted as a Gibbs-Duhem relation \cite{Gibbons}.
Indeed lets suppose we  send a particle of momenta $p_{\mu}$ inside the screen,
 its energy is given by
$e= - \bm p\cdot \bht$ and its angular momenta by $j\cdot \bm \varphi= \bm p \cdot \bm\varphi$.
 A momenta is flowing out of the bulk region if $0 \leq -\bm p \cdot \t = e-j\cdot \bm \varphi $.
 In this case, and as we will see in full generality later, under some equilibrium  conditions\footnote{We will precisely identify these conditions as the preservation of the internal energy of the screen  together with the condition that the process do not generate any gravitational waves.}, we have  that
 the energy-momenta flow can be registered on the screen by an increase in area:
 \be
 0 \leq e - j\cdot \bm \varphi = T \delta S 
 \ee
 Thus the change in area is given by a first law,
 \be\label{diffe}
 \delta M = T\delta S + \Omega \delta J,
 \ee
 where $\delta M=e$ $\delta J = j_{\bm r}$.
 Both $M$, $S$ and $J$ are extensive variables homogeneous under rescaling of length.
 However $M$ is homogeneous of order $1$ while $A$ and $J$ are homogeneous of order $2$.
 Using this we can easily integrate out \Ref{diffe}, assuming there is no residual contribution at zero size,  into the generalized Gibbs-Duhem relation \Ref{GD}.
 In other word the factor $1/2$ entering the relationship between the mass and the canonical energy, is not an anomaly but the reflexion of the thermodynamical nature of the relationship.

\subsection{Canonical energy: Gibbs energy versus internal energy}
 
 In the previous sections we have constructed the canonical energy associated with the Einstein Lagrangian and decomposed this energy in terms of a screen and pure  boundary terms.
 This energy is uniquely defined once we chose which Lagrangian we work with. 
 We can, however, add a boundary term to the gravity Lagrangian, which will define a new type of energy.
 This should not be surprising since it is also the case in thermodynamic that the energy depends on what is kept fixed, 
 and different energies are related by canonical transformations. For instance, we can talk about the internal energy 
or the  free energy or Gibbs energy. They are all related by Legendre transforms which change which quantities are kept fixed.
 The internal energy for a closed system $U(S,V)$ is characterized by the fact that it depends on extensive quantities, $\rd U = T\rd S -p\rd V$. That is quantities like $V$ and $S$ that scale with the size of the system.
 While on the other end of the spectra the Gibbs energy   $G(T,p)=U-TS+pV$ depends on intensive quantities, $\rd G= V\rd p - S\rd T$, 
 that do not scale with the size of the system.
 
 We have seen that the on-shell variation of the Einstein Lagrangian leads to a boundary term 
 $\nabla_{\mu}\alpha^{\mu}$ where $\alpha^{\mu}$ is defined in terms of the variation 
 of the connection $\delta \Gamma_{\a\b}^{\nu}$ on the boundary. The connection coefficient are invariant under  space-time independent rescaling\footnote{Since the Einstein tensor is invariant under such rescaling we can consider that this variation is an on-shell variation, once we rescale the matter field appropriately.} of the metric $\delta g_{\a\b}= \phi g_{\a\b}$.
 So we can think of the component of the connection on the boundary as the intensive variables.
 This means that the canonical energy and its screen component $G_{\t}$ that we have just defined is the gravitational analog of the Gibbs energy. 
 
 On the other hand the metric components on the boundary are  analogous to extensive variables, since they rescale homogeneously. It is well-known that the addition of the Gibbons-Hawking boundary term \cite{Gibbons-H}
 leads to an on-shell variation which is proportional to variation of the boundary metric. The canonical energy derived from the Hawking-Gibbons modification of Einstein Lagrangian is therefore  the analog of the internal energy. We are going to see that this internal energy coincides with the Brown-and York energy \cite{Var3}.
 
 Lets  now investigate what happens when we modify the Einstein Lagrangian by a boundary term 
 \be
 \cL' = \cL + \N_{\mu} V^{\mu},
 \ee
 which corresponds to a modification of the action by boundary contribution
 \be
 S' = S - \int_{\Sigma_{i}}^{\Sigma_{f}} \sqrt{h} V_{\bbn} + \int_{\bar\Sigma_{i}}^{\bar\Sigma_{o}} \sqrt{\bar{h}} V_{\s}.
 \ee
 where $\Sigma_{f}$ (reps. $\Sigma_{i}$) denotes the final (reps. initial) slices, 
 $\bar\Sigma_{o}$ (reps. $\bar\Sigma_{i}$) denotes the outer (reps. innner) boundaries,
 and $\bbn$ is directed forward in time while $\s$ is directed outward.
 The symplectic potential becomes 
$
 \alpha'^{\mu}= \alpha^{\mu} + \delta V^{\mu} + V^{\mu} \frac12 g^{\a\b}\delta g_{\a\b}
$, thus 
%\be
%\alpha'^\mu n_{\mu} = \alpha^{\mu}n_{\mu} + \delta V_{\n} + V_{\n} \frac12 g^{\a\b}\delta h_{\a\b} + V^{\mu} h_{\mu}{}^{\nu} \delta n_{\nu}.
%\ee
the variation of the canonical energy is therefore given by 
$$H'_{\t} = H_{\t} +\int_{\Sigma}(\cL_{\t} V_{\n} + h^{\a\b}\N_{\a}t_{\b} V_{\n} - N \N_{\a}V^{\a})\sqrt{h}$$
 where we have used  that $h_{\a}{}^{\b} \cL_{\t} n_{\b}=0$ and we denote $V_{\n}=
 V^{\mu}n_{\mu}$.
 Given the decomposition of the time vector as $\t = N \n + M\s +\bm\varphi$ we can show that the additional term is a total derivative\footnote{ we use that $\cL_{N\n}V_{\n} +h^{\a\b}\N_{\a}(N\n_{\b}) V_{\n} = N\N_{\mu}( n^{\mu} V^{\n})$ and that 
 $N \N_{\mu}( h^{\mu}{}_{\nu} V^{\nu}) = D_{a}(Nh^{a}{}_{\mu}V^{\mu})$.}:
 $ D_{\a}(N h^{\a}{}_{\b} V^{\b} + (M s^{\a}+\varphi^{\a}) V_{\n})$.
 Integrating by part and making use of $NV_{\s}+MV_{\n}= \rho V_{\bbs}$
 where $\rho$ is the boundary lapse and $\bbs$ the space like vector normal to $\bar{\Sigma}$
 directed towards the outer region.
We get  the canonical energy associated with  the new Lagrangian to be:
 \be
 H'_{\t} = H_{\t} - \int_{ S_{i}}^{S_{o}} \sqrt{q}  \rho V_{\bbs}.
 \ee
 If the canonical energy of the  Einstein Lagrangian is the Gibbs energy, 
 the analog of the gravitational internal  energy is the energy obtained by Legendre transform of the  canonical energy $H$. It is obtained by adding to the Einstein Lagrangian the Gibbons Hawking boundary term: $$ V^{\mu}_{GH} \equiv  \frac1{8\pi G \cosh \beta}\left(\bar{n}^{\mu} (\N_{\a}n^{\a}) + {s}^{\mu } (\N_{\a}\bar{s}^{\a}) \right).$$
 This term is characterize by the property that $(8\pi G)V_{\n}^{GH} = - (\N_{\a}n^{\a})$ and 
 $(8\pi G)V_{\bbs}^{GH}= (\N_{\a}\bs^{\a})$ where $\n$ is the timelike normal to  the slice $\Sigma$ and $\bbs$ is the space like normal to the screen, while $ K=(\N_{\a}n^{\a})$ and $\bar{H}=(\N_{\a}\bs^{\a}) $ represent the trace of the extrinsic curvature of $\Sigma$ and $\bar{\Sigma}$ respectively. 
 The internal energy denoted $U_{\t}$ is therefore given by
 \be
 U_{\t} = G_{\t} - \int_{S} \sqrt{q} \rho \frac{(\N_{\a}\bs^{\a})}{8\pi G}.
 \ee
 where $\rho \equiv N/\cosh \beta$ is the boundary lapse introduced earlier. 
 In order to evaluate this we expand in terms of the 2d variables both the addition term 
 $\N_{\a}\bs^{\a} = \bbs\dd\N_{\bbn}\bbn + \theta_{\bbs}$ and the Gibbs energy term
  %= \frac12(\bbs\dd\N_{\bbn}\bbn + \N_{\bbs}\rho)
  $\bar\gamma_{\t}= \rho \bbs\dd\N_{\bbn}\bbn  -(8\pi G) j_{\bm \varphi}.$
  Here $\theta_{\bbs}\equiv (\cL_{\bbs}\sqrt{q})/\sqrt{q}$ is the 2-dimensional extrinsic curvature of the screen.
 Thus 
 \be
 U_{\t}=- \int_{S} \sqrt{q}  \left(\frac{\theta_{\bot}}{8\pi G}+ j_{\bm \varphi} \right)
 % - \frac{1}{16\pi G}\int_{S} \sqrt{q} \cL_{\t}\bar{\beta}  
 \ee
 where $\bot\equiv -(\t\dd\n) \s +(\t\dd\s)\n
$ is the radial vector normal to the screen and $\bht$.
This shows that we can associate to the screen an internal energy density given by
 \be
 \boxed{
 \epsilon \equiv -\frac{\theta_{\bot}}{8\pi G}.
 }
 \ee
 This accounts for the modification of the screen energy.
 
 We can also perform a canonical transformation that affects the boundary Gibbs energy $h_{\t}$.
 In order to do so we need to modify the actions at the corners $S =\Sigma\cap\bar{\Sigma}$ of the boundary:
Lets suppose  that the addition to the Lagrangian is of the form $\nabla_{\mu}V^{\mu}$ where 
$V$ is itself a boundary contribution:
\be
V^{\mu} = \f1{\cosh \beta}(\bn^{\mu} D_{a}(v^{a}_{\n}) + s^{\mu} \bar{D}_{a}(\bar{v}^{a}_{\bbs})).
%=\f1{\cosh \beta}(\bn^{\mu} n_{\a} \N_{\b} W^{\a\b} - s^{\mu} \bs_{\a}\N_{\b} \bar{W}^{\a\b})
\ee
This corresponds to a modification of the Lagrangian at the two sphere $S=\Sigma \cap \bar{\Sigma}$ of the form
\be
\delta S =  \int_{\Sigma_{i}}^{\Sigma_{f}} \sqrt{h} D_{a}(v^{a}_{\n}) + \int_{\bar\Sigma_{i}}^{\bar\Sigma_{o}} \sqrt{\bar{h}} \bar{D}_{a}(\bar{v}^{a}_{\bbs})= \int_{S_{fi}\cup S_{io} }^{S_{fo}\cup S_{ii}}\sqrt{q}(v_{\s\n}-\bar{v}_{\bbn\bbs}).
\ee
where $v_{\s\n}\equiv s_{a}\d_{\n}^{a}$.
On the other hand for a closed surface $S$
 it modifies the Hamiltonian  by a total  time derivative\footnote{ We use 
that \bea
\rho \bar{D}_{b}({\d}^{b}_{\bbs}) = 
-\rho \bar{D}_{b}(\bn^{b}{\d}_{\bbn\bbs}) +\rho \bar{D}_{c}(q^{c}{}_{b}{\d}^{b}_{\bbs})
= -\rho(\cL_{\bbn} + \theta_{\bbn}){\d}_{\bbn\bbs} + \bm{\rd}\dd(\rho\bm q\dd {\d}_{\bbs})
\eea together with $\t =\rho \bbn +\bphi$ and $ \d_{\bbs\bbn}=-\d_{\bbn\bbs}$.}
\be
\delta H 
%= - \int_{S} \sqrt{q} \rho \overline{D}_{\b}(\bar{v}^{\b}_{\bbs})
%=  \int_{S} \cL_{\rho \bbn}\left(\sqrt{q} \bar{v}_{\bbn\bbs}  \right) 
= \int_{S} \cL_{\t}\left(\sqrt{q} \overline{v}_{\bbn\bbs}  \right).
\ee
%In the case at hand if one take
%$
%v_{\n}^{a}\equiv s^{a} \eta$, and $
%\bar{v}_{\bbs}^{a}\equiv -\bbn \bar{\eta}$, 
%we get that the modification of the action and the new Hamiltonian are given by
%\be\d S=   -\int_{S}\sqrt{q} \frac{\beta}{8\pi G},\qquad 
%h'_{\t}= \frac1{16\pi G} \int_{S} (\cL_{\t}\sqrt{q}).
%\ee
%\be
%v_{\s\n}- \bar{v}_{\bbn\bbs}= \frac{\eta - \bar\eta}{16\pi G} = - \frac{\beta}{8\pi G}.
%\ee
%is the modification to the action, while the new boundary Hamiltonian is 
%\be
%h'_{\t}= \frac1{16\pi G} \int_{S} (\cL_{\t}\sqrt{q}) \eta.
%\ee 
%\cite{Hawking-Hunter}
%
%
%

\section{2+2 decomposition}\label{22}

The system we are interested in is a two sphere $S$ which lies at the intersection of a timelike screen $\bS$ and a spacelike  surface $\Sigma$.
We denote by $\n$ the unit timelike vector normal to 
$\Sigma$ and by $\s$ the unit spacelike vector tangent to $\Sigma$ and normal to $S$.
Similarly we denote by $\bbs$ the unit spacelike normal to $\bS$ and by $\bbn$ the unit timelike vector tangent to $\bS$ and normal to $S$. We are interested in a region of space-time that extend to the future of $S$ and outside of the screen $\bS$.
We assume that in the region of space-time that we are interested in there is a double foliation, where  $\Sigma_{t}$ are the spacelike leaves of a foliation given by $T=t$, while the screens 
$\bS_{r}$ are the timelike leaves of a foliation given by $ R=r$; and such that $S_{t,r} =\Sigma_{t}\cup \Sigma_{r}$ is a 2-sphere. 
We assume that $T$ increase in the future and that $R$ increase when we go away from the screen.

As we have seen this double foliation is characterized by $3$ scalars $ \rho, \tau, \beta$.
The first scalar that characterizes the double foliation is the boost angle $\beta$. This corresponds to the boost 
velocity that a  screen observer following the tangent $\bbn$ possesses compare to an Eulerian observer 
$\n$ at rest with respect to the foliation $\Sigma_{t}$:
\be
\bbn =  \cosh \beta \n +  \sinh \beta \s, \qquad \bbs = \cosh \beta \s + \sinh \beta \n.
\ee 
For the other scalars,
it is clear that $\rd T$ is proportional to the one form $n_{\mu}\rd x^{\mu}$ and that 
$\rd R$ is proportional to the one form $\bs_{\mu} \rd x^{\mu}$. $\rho$ and $\tau$ characterizes the proportionality coefficients
\be\label{defns}
\n = - \rho\cosh \beta  \bm\N T,\qquad \bbs = \tau\cosh \beta \bm\N R.
\ee
The signs are chosen such that the vector  $n^{\mu}$ is future directed and $\bs^{\mu}$ outwardly. The factors of $\cosh \beta$ are chosen such that  if one defines
\be
\bht \equiv \rho \bbn,\quad \quad  \bhr\equiv  \tau \s.
\ee
Then  $\bht$ is an evolution vector tangent to the screen $\bS_{r}$
which is Lie dragging $\Sigma_{t}$ (i-e $ \cL_{\bht} T =1$).
Similarly,  the  displacement vector $\bhr$   is tangent to the initial data surface $\Sigma_{t}$
which is Lie dragging $\bS_{r}$ (i-e $ \cL_{\bhr} R =1$).
These vectors also preserve the double foliation since
\be
\cL_{\bht} R =0,\qquad \cL_{\bhr} T =0.
\ee
We cannot in general represent these vectors as velocity vectors $ \hat{t}^{\mu} =\frac{\partial x^{\mu}}{\partial T}$ and 
$ \hat{r}^{\mu} =\frac{\partial x^{\mu}}{\partial R}$, unless  the two vector fields commutes.
The properties of hyper surface orthogonality, i-e $\cL_{\bht} T =\cL_{\bhr} R =1$, of the vectors $\bht,\bhr$ means that
$$ s^{a}\cL_{\bht} n_{a}=0,\qquad  \bn^{a}\cL_{\bhr} \bs_{a}=0.$$
This implies that their commutators is proportional to the the twist vector that measure the anholonomicity (non integrability) of the two planes $TS^{\perp}$, \be [\bht,\bhr] =  (\rho \tau\cosh \beta)\,  \bmo, \mrm{with} \omega_{A} \equiv  [n,s]^{\a}q_{\a A}.\ee
By Frobenius theorem, the vanishing of $\bmo$ implies that the normal two planes $TS^{\perp}$ are integrable, i-e they can be understood as the tangent vectors to a submanifold. 
In general the vectors $\bht,\bhr$ do not commute,
this means that  we need to define time flow vectors $\t \equiv \partial_{T}$ and $ \bm{r} \equiv \partial_{R}$ that Lie-commute:
\be
\t \equiv \bht + \bm\varphi,\qquad \r \equiv \bhr +\bm \psi,
\ee
where  $\bm\varphi$ and $\bm \psi$ are vectors tangent to $S$ and
 $\bm\varphi$ is the angular velocity. This vectors Lie-commute if $\bm\psi $ is chosen in order to satisfy 
$$\partial_{R} \bm\varphi -\partial_{T} \bm\psi + [\bm\varphi,\bm\psi]=   (\rho \tau\cosh \beta)\bmo.$$
Note that if we interpret $(\bm\varphi,\bm \psi)$ to be  a 2d gauge field valued into the algebra of 2d diffeomorphisms.
The LHS of the previous equation is simply the curvature of this gauge field since the commutator is given by the Lie bracket.  The previous equality expresses that this curvature is proportional to the twist. 

We can now express the space-time metric and its inverse, in terms of the normal coordinates $(T,R)$ the sphere coordinates $x^{A}$, the foliation scalars $(\rho,\beta,\tau)$ and the velocity vectors $(\bm{\varphi},\bm \psi)$ as follows
\bea
\rd s^{2} =  \left(-\rho^{2}\rd T^{2} + 2 (\rho \tau \sinh \beta)\, \rd T \rd R + \tau^{2} \rd R^{2} \right)
%+ q_{AB} \left(\rd x^{A} + \varphi^{A} \rd T + \psi^{A} \rd R \right)
%\left(\rd x^{B} + \varphi^{B} \rd T + \psi^{B} \rd R \right)\\
+ \left(\rd \bm{x} + \bm\varphi \rd T + \bm\psi \rd R \right)\dd \bm q \dd
\left(\rd \bm{x} + \bm\varphi \rd T + \bm\psi \rd R \right).
\eea
This metric can be inverted and we can read the normal components of the inverse metric to be 
\be
g^{TT} = - \frac1{\rho^{2}\cosh^{2}\beta},\qquad g^{TR}=  \frac{\sinh \beta} {\rho\tau\cosh^{2}\beta},
\qquad
g^{RR} = \frac1{\tau^{2}\cosh^{2}\beta}.
\ee
This is 
in agreement with \Ref{defns}, since we can check that $g(\n,\n)=-1$, $g(\bbs,\bbs)=1$ and $g(\bbs,\n)=\sinh\beta$.
The off diagonal components are 
\be
g^{AT} 
%= - g^{TT}\varphi^{A}-g^{TR}\psi^{A}
=\frac{\tau \varphi^{A}- \rho\sinh \beta \psi^{A}}{\rho^{2}\tau\cosh^{2}\beta}
,\qquad 
g^{AR}
%= -g^{RR}\psi^{A}-g^{RT}\varphi^{A}
= -\frac{\rho \psi^{A}+\sinh \beta\tau  \varphi^{A}}{\rho\tau^{2}\cosh^{2}\beta},
\ee
while the tangential components are
$
g^{AB} = q^{AB } + g^{TT}\varphi^A\varphi^B + 2 g^{TR}\varphi^{(A} \psi^{B)} + g^{RR}\psi^{A}\psi^{B}.
$
If one restricts the space-time metric to a spacelike or timelike slice,
one sees that $(\rho, \bm\varphi) $ are the lapse and shift of the induced metric on 
the timelike screen $\bar{\Sigma}_{r}$;
 while $(\tau,\bm\psi)$ is the corresponding ``spatial'' lapse and shift on the slice $\Sigma_{t}$:
\bea
\rd s^{2}_{\Sigma} &=&  -\rho^{2}\rd T^{2}
+ \left(\rd \bm{x} + \bm\varphi \rd T \right)\dd \bm q \dd
\left(\rd \bm{x} + \bm\varphi \rd T  \right),\\
\rd s^{2}_{\bar\Sigma} &=&   \tau^{2} \rd R^{2} 
+ \left(\rd \bm{x} + \bm\psi \rd R \right)\dd \bm q \dd
\left(\rd \bm{x}  + \bm\psi \rd R \right),
%\\
%\rd s^{2}_{S} &=&  
%\rd \bm{x} \dd \bm q \dd
%\rd \bm{x} .
\eea
Once we know the metrics on the timelike slices $\bar\Sigma_{r}$ and the spacelike slices $\Sigma_{t}$, and assuming that  their pullback agree on $S$, we can reconstruct the space-time metric provided we know the value of $\beta$.

\subsection{Intrinsic and extrinsic geometry}\label{int-ext}

The geometry of the embedding of the 2d sphere $S$ in the space-time  is characterized by 
two types of geometry: intrinsic, extrinsic and by the accelerations.

The intrinsic geometry is determined by the 2d metric $q_{a b } = g_{ab } + n_{a}n_{b} - s_{a}s_{b}$.
By construction we have that $q_{ab} n^{b} = q_{ab} s^{b}=0$.
In the following we will denote by an uppercase indices the projection of space-time vectors onto vectors tangent to $S$: 
$v_{A} \equiv q_{Ab } v^{b}$. The Levi-Civita connection associated with $q$ is denoted by $\rd_{A}$; it acts on vectors tangent to $S$ as $ \rd_{A} v^{B} = q^{B}{}_{b} q_{A}{}^{a} \N_{a} v^{b}= q^{B}{}_{b}  \N_{A} v^{b}$.
The extrinsic geometry of $S$ is characterized by a deformation tensor  $\Theta^{\bm\ell}_{AB}$ for any vector $\bm\ell$ normal to $S$:
\be
\Theta_{\bm\ell AB} = \N_{A} \ell_{B} \equiv q_{A}^{a}q_{B}^{b} \N_{a}\ell_{b}.
\ee
From this definition it follows  that $\bm \Theta_{\bm \ell}$ is a symmetric tensor and that it it linear in its argument: $\bm{\Theta}_{a\n+b\s}= a \bm{\Theta}_{\n}+b \bm\Theta_{\s}$.
We denote its trace $\theta_{\bm\ell}$ and 
in the following we will also used the notation
\be
\theta_{\bm\ell} \equiv q^{AB}\Theta_{\bm\ell AB },\qquad  \widehat{\Theta}^{AB}_{\bm \ell}\equiv {\Theta}^{AB}_{\bm \ell} - {q^{AB}} \theta_{\bm \ell}.
\ee

Since we have two normal directions, the deformation tensor do not fully characterize the extrinsic geometry,
we also have the {\it normal connection} $\bm \pi$ which is the projection along $S$  of  $ q_{A}{}^{a} \s\dd \N_{a} \n$:
\be
\pi_{A} \equiv q_{A}{}^{a} s_{b}\N_{a} n^b = \s\dd \N_{A} \n.
\ee
This form depends on the choice of a basis  $(\n,\s)$ of the normal direction space. If one chose the basis adapted to the screen instead we have the relationship
\be
\bar{\pi}_{A} \equiv  \bbs\dd \N_{A} \bbn = \pi_{A} +\rd_{A}\beta.
\ee
To summarize, the data $(\bm{\Theta}_{\n},\bm\Theta_{\s},\bm\pi)$ characterizes the extrinsic geometry.

The final data is associated with the accelerations.
First, we have the normal accelerations 
\be
\gamma_{\n}\equiv  \s\dd \N_{\n} \n,\qquad  \gamma_{\s}\equiv  -\n\dd \N_{\s} \s.
\ee
The first one is the radial acceleration of Eulerian observers, static with respect to the foliation $\Sigma_{t}$;
together with their screens analog. The change of frame amounts to the gauge transformation: $\bar\gamma_{\bm\ell}=\gamma_{\bm\ell} +\N_{\bm\ell}\beta$.

We also have the tangential accelerations 
which are of different types: 
First the timelike and space like  accelerations:
$$a^{A}_{\n}\equiv q^{A}{}_{a} \N_{\n}\n^{a},\qquad {a}^{A}_{\s}\equiv - q^{A}{}_{a} \N_{\s}\s^{a}.$$
$a^{A}_{\n}$ is the acceleration of the Eulerian observers at rest with respect to the foliation,
while ${a}^{A}_{\s}$ is the acceleration  observers following a radial evolution normal to the screens.
We also introduce the screens analog denoted $\bm{\bar{a}}_{\bbn}, \bm{\bar{a}}_{\bbs}$.
%
%$\bar{a}^{A}_{\bbn}\equiv q^{A}{}_{a} \N_{\bbn}\bbn^{a}$ is the acceleration of the  observers in the screen with a velocity normal to the foliation.
%Similarly, we have the space like accelerations $\bar{a}^{A}_{\bbs}= - q^{A}{}_{a} \N_{\bbs}\bbs^{a}$ of observer at a fixed moment of time on the screen and ${a}^{A}_{\s}= - q^{A}{}_{a} \N_{\s}\s^{a}$ of the observers following a radial evolution normal to the screens.
These four accelerations are not independent since the total acceleration is independent of the frame 
$\aa_{\n}+\aa_{\s}= \bar{\aa}_{\bbn}+\bar{\aa}_{\bbs}$.
The three independent components can be written in a frame as the frame independent total acceleration $\aa_{\n}+\aa_{\s}$ and the two frame dependent components
$\aa_{\n}-\aa_{\s}$ and $2\bm{b}\equiv \bm q\cdot(\N_{\s}\n +\N_{\n}\s)$.

The last data is the twist vector that we already have introduced it is given by
\be
\bm\omega = \N_{\n}\s -\N_{\s}\n.
\ee
It is independent of the frame.

All the accelerations can be expressed in terms of the foliation scalars $\rho,\beta, \tau$.
The detail derivation of these relationships is given in appendix \ref{acc-sec},
 we just  sketch this derivation here. One first expresses that  the time flow vector $\bht =\rho\bbn$ 
preserves the foliation $\Sigma_{t}$ and the space translation vector $\bhr=\tau\s$ preserves the foliations
$\bS_{r}$ which implies that $ s^{b}\cL_{\bht} n_{a}=0$ and 
$\bn^{b}\cL_{\bhr} \bs_{a}=0$.
These conditions entirely determine the normal accelerations in terms of the foliation scalars.
One finds 
\be\label{acc0}
\gamma_{\n} = \frac{\N_{\s}(\rho \cosh \beta)}{\rho \cosh \beta},\qquad
\bar{\gamma}_{\bbs}= \frac{\N_{\bbn}(\tau \cosh \beta)}{\tau \cosh \beta}.
%\gamma_{\s} =\frac{\N_{\bht}\tau - \N_{\bhr}(\rho \sinh \beta)}{\rho\tau \cosh \beta},
%\gamma_{\bhr} =\frac{\N_{\bht}\tau -  \sinh \beta\N_{\bhr}\rho}{\rho \cosh \beta} - \N_{\bhr} \beta,
\ee
Now using  that  $ \bbn = \frac{\n}{\cosh\beta} + \tanh \beta \bbs$, and $\s = \frac{\bbs}{\cosh\beta} - \tanh \beta \n$ and that $\bar{\gamma}_{\bm \ell}=\gamma +\N_{\bm\ell}\beta$ we can obtain the other components, for instance  that
\be\label{gamman}
\bar{\gamma}_{\bbn}= \frac{\N_{\bbs}\rho}{\rho} + \N_{\bbn}\beta + \tanh\beta\left(\frac{\N_{\bbn}\tau}{\tau}-\frac{\N_{\bbn}\rho}{\rho}\right).
%\qquad
%\gamma_{\s}= \frac{\N_{\n}\tau}{\tau} - \N_{\s}\beta + \tanh\beta\left(\frac{\N_{\s}\tau}{\tau}-\frac{\N_{\s}\rho}{\rho}\right).
\ee
A detail derivation is provided in the appendix.

We can also express the tangential accelerations in terms of the foliation scalars using that the double foliation flow vectors $\bht,\bhr$ preserves the double foliation,
that is $q_{a}{}^{b}\cL_{\bht} \bn_{a} = 0= q_{a}{}^{b}\cL_{\bht} \bs_{a}$,  while 
$q_{a}{}^{b}\cL_{\bhr} s_{a}=0=q_{a}{}^{b}\cL_{\bhr} n_{a}$.
This implies that 
\be\label{acc1}
 {\bm a}_{\s}
= \frac{\bm\rd \tau}{\tau },\qquad {\bm a}_{\n}
= \frac{\bm\rd (\rho \cosh \beta)}{\rho \cosh \beta},\qquad 
2\bm{b} =   \bm\rd \beta+ \tanh \beta \left(\frac{\bm\rd \rho}{\rho}-\frac{\bm\rd \tau}{\tau}   \right) .
\ee
where $\bm{b}=\frac12(\N_{\s}\n+\N_{\s}\n)$.
The last object we need to determine is the twist vector $\bm{\omega} = (\N_{\n}\s -\N_{\s}\n)$ which depends on the normal form and the foliation scalars as 
\be
2\bm{\omega}= -2\bm \pi  +  \bm\rd \beta +  \tanh \beta \left(\frac{\bm\rd \rho}{\rho}-\frac{\bm\rd \tau}{\tau}   \right). 
\ee

In summary, we have shown that the double foliation is characterized by
 three scalars : The timelike lapse $\rho$, the space like lapse $\tau$, the boost parameters $\beta$, which determine the choice of foliation.
  While the geometry of this foliation is given by the intrinsic 2d metric $\bm q$, the extrinsic deformation tensors $(\bm\Theta_{\n}, \bm \Theta_{\s})$,
and the normal connection $\bm \pi$. 
The hyper surface orthogonality allow to express, according to (\ref{acc0}, \ref{gamman}, \ref{acc1}),  all the accelerations in terms of the foliation scalars. These 
data are independent of the choice of frame $(\n,\s)$ except the normal connection since $\bm \bpi =\bm \pi+\bm \rd \beta$.
Finally the twist one form $\bm \omega$ which determines to what extent the normal 2-planes are integrable is characterized by the normal connection and the foliation scalars.

\subsection{The many faces of acceleration}\label{sec-manyacc}

According to section \ref{Cenergy} the surface energy density is given by a sum of two terms:
$
\kappa_{\t} = \gamma_{\t} - \frac12\delta_{\t}
$
where 
\be
\gamma_{\t}\equiv \s \N_{\t} \n,\qquad \delta_{\t} =\left( s_{\mu}\cL_{\t}n^{\mu}+s^{\mu}\cL_{\t}n_{\mu}\right).
\ee
%In the expression for $\delta_{\t}$ we have used the hypersurface orthogonality of the time evolution: $s^{\mu}\cL_{\t}n_{\mu}= 0$.
%=  \s \cdot \N_{\t} \n  +  \n\cdot \nabla_{\s}\t $. 
%This implies as we have seen in the previous section that
%\be\label{hyport}
%\omega_{\n} = \frac{\nabla_{\s}\rho}{\rho} +\tanh\eta \N_{\s}\eta.
%\ee
%
In order to understand the nature of the two terms appearing in the definition of the surface energy density,
let us remark that each term $\gamma_{\t}$ and $\kappa_{\t}$ are  antisymmetric in the 
exchange of $\n$ and $\s$.
They are not, however, individually invariant under boost transformations (or change of frame) defined by $\delta_{\beta} \n = \beta \s$,  $\delta_{\beta} \s = \beta \n$. Indeed, under such transformations we have $ \delta_{\beta} \gamma_{\t} =\N_{\t}\beta$, and 
$ \delta_{\beta} \delta_{\t} =2 \N_{\t}\beta$. This implies however that $\kappa_{\t}$ is invariant under boosts  hence under change of frames.

This frame independence can be rendered manifest by expanding the Lie derivatives in the definition of $\delta_{\t}$. Using that $s^{\mu}\cL_{\t}n_{\mu}= \s \cdot \N_{\t} \n  +  \n\cdot \nabla_{\s}\t $ and that 
$s_{\mu}\cL_{\t}n^{\mu}= 0=  \s \cdot \N_{\t} \n  -  \s\cdot \nabla_{\n}\t $ we easily get 
\be\label{acc3}
\boxed{\kappa_{\t} = \f12\left(  \s\cdot \N_{\n}\t- \n\cdot \N_{\s} \t \right)  =  n_{\mu} s_{\nu}  \nabla^{[\mu}t^{\nu]}.
%= \t \cdot [\s,\n]
}
\ee
This expression makes it manifest that  it is boost invariant, it doesn't depend on the foliation but only on the normal subspace to $S_{t}$ spanned by the bivector $\n\wedge \s$.
It is also clear from this expression that it coincides with the Komar expression, in the case $\t$ is a killing field \cite{Komar}.

Let us now use the decomposition of the time flow in terms of its normal and tangential components.
That is we define 
\be
\bht = \rho \bbn, \quad \bbn\equiv \cosh \beta \n + \sinh \beta \s,\quad \mrm{and} \quad \t =\bht +\bm{\varphi}.\label{dt}
\ee
$\bm \varphi$ is the tangential component of $\t$,
also  $\rho^{2} = - \bht \cdot \bht$ is the time lapse of the double foliation introduced earlier  and  $\beta$ is the boost angle between the screen and the foliation, i-e
 $\rho\sinh \beta \equiv  \t\cdot \s$.
We also introduce
\be
{\bot} \equiv \rho \bbs,\qquad \bbs\equiv \cosh \beta \s + \sinh \beta \n,
\ee
where $\bbs$ is a unit spacelike vector perpendicular to $\bht$ and to $S_{t}$.

From the definition \Ref{acc3} and the fact that $\bm \varphi$ is tangent to $S$ we can establish that 
\be
\kappa_{\t} =\kappa_{\bht} - \bm\varphi \cdot \frac12 [\n,\s] =\kappa_{\bht} - \frac12 \bm\varphi \cdot \bm \omega .
\ee
where $\bm \omega$ is the twist one form. In other words we have that the total mass density 
and angular momenta
density  are given by $\rho_{\t}=\kappa_{\bht}/4\pi G$,and $j_{\bphi}=-\kappa_{\bphi}/8\pi G$:
\be
\rho_{\t}\equiv \frac{(\s\dd\N_{\n}\bht - \n\dd \N_{\s}\bht)}{8\pi G},\qquad j_{\bm \varphi} =  \frac{\bm\varphi \cdot \bm \omega}{16\pi G}.
\label{rhoj}
\ee
As we have seen these quantities can be decomposed in terms of a  screen surface tension $\sigma_{\t}$ and momenta $\pi_{\bphi}$ given by
\be
\sigma_{\t}=- \frac{\bar\gamma_{\bht}}{8\pi G},\qquad p_{\bphi}= \frac{\bm\pi\dd\bphi}{8\pi G},
\ee
and a boundary contribution proportional to $-\bar\delta_{\t}/(8\pi G)=\cL_{\t}\bar\eta$, which is interpreted as the boost energy. 
Indeed we have that 
\be
\rho_{\t}= -2\sigma_{\t} -\frac{\bar\delta_{\bht}}{8\pi G},\qquad j_{\bm \varphi} =p_{\bphi} + \frac{\bar\delta_{\bphi}}{16\pi G}.
\ee These contributions depends not only on the screen but also on the way the slicing is boosted relatively to the screen. The expression for the boost energy can be given in terms of the foliation scalars
as (see \Ref{dvar} and appendix \ref{acc-sec})
\be
\bar{\d}_{\t}=  \cL_{\t} \beta + {\tanh \beta} \left( \frac{\cL_{\t} \tau}{\tau} - \frac{\cL_{\t} \rho}{\rho} \right)\equiv \cL_{\t}\eta.
%=\frac{1}{2 \cosh \beta} 
%\cL_{\bht}  \ln\left(\frac{\tau}{\rho} \sinh \beta \right)
\ee

In order to understand the meaning of the formula for the mass density $\rho_{\t}$ and the surface tension $\sigma_{\t}$ let us start to evaluate it for a particular  bulk foliation orthogonal to the screen, that is where where  $\bht = \rho \n $. 
This expresses that the boundary observers on the screen are static with respect to the foliation.
In other words the boundary observers  coincide with the fiducial observers.
In this case we have that  $\delta_{\t}=0$ the boost energy vanish and $\rho_{t}= -2 \sigma_{t}$.
 This express the fact that the observer is purely sliding along the screen. The energy surface density is then just equal to $\rho_{\t} =  \s\cdot \bm{a}_{\bht}/ 4\pi G $ which is the projection of the acceleration $\bm{a}_{\bht}=\N_{\bht}\n$ of the fiducial observers along the radial direction.
So in this case $\rho_{\t}$ is simply the radial acceleration in planck unit:
\be\label{sigma1}
\rho_{\t} =   \frac{\bm{a}_{\t} \cdot\s}{4\pi G}  = \frac{ \N_{\bot}\phi}{4\pi G},
\ee
where  we have introduced the {\it Newtonian potential}
$ \phi\equiv \ln \rho $ associated with the foliation $\Sigma$ and 
$\N_{\bot} \phi = \N_{\bbs} \rho$  corresponds to the acceleration of screen observers with respect to static fiducial observers\footnote{ In the case of a static Schwarzchild black Hole we have that 
$\phi(r) =\frac12 \ln\left(1-\frac{2GM}{r}\right)\simeq -\frac{GM}{r}$ which is indeed the Newtonian potential. Moreover we have that $ \bot =\left(1-\frac{2GM}{r}\right) \partial_{r}$ therefore one gets that for a screen at a distance $R$ in a Schwarzchild space time the energy density  is 
$$\rho_{\t} = \left(1-\frac{2GM}{R}\right) \partial_{R}\phi(R) = \frac{GM}{R^{2}}. 
$$ The mass is then expressed as $M=\frac{\sigma A}{4\pi G}$.}. The equality \Ref{sigma1} follows from the hyper surface orthogonality condition \Ref{acc0} or (\ref{gamman}) when $\beta=0$.

Far away from the sources of gravitation, $\phi $ is  the Newtonian potential, therefore $\rho_{\t}$ is the acceleration needed for an observer to stay static.
This  is {\it  positive} when the acceleration is directed toward the {\it inside  } of the screen.
That is it is positive for an outer screen. Similarly the surface tension $\sigma_{\bht}=-\rho_{\t}/2$ is generally positive for an inner screen.

It is illuminating to notice that the formula for the mass that we have given is similar to the one we obtain in the Newtonian regime. Indeed in newtonian gravity the Poisson equation gives us $\Delta \varphi = 4 \pi G \rho$ where $\rho$ is the matter energy density and $\varphi$ is the Newtonian potential.
The total newtonian mass in a volume $V$ can then be evaluated by a Gauss law formula as 
$$
M_{Newt}= \int_{S} \sqrt{q} \rho_{Newt}, \qquad \rho_{Newt}= \frac{\nabla_{r}\varphi}{4\pi G} 
= \frac{a_{r}}{4\pi G},
$$
where $\partial V =S$ and the last equality expresses that  the radial inward acceleration $a_{r}$ is equal to $\N_{r}\varphi$ by Newton's law. The formula for the mass we have is therefore a covariant generalization of the Newtonian Gauss law.

Let us now deal with the general case. The surface tension is always proportional to the inward radial acceleration. We need then to understand the meaning of the difference $\bar{\delta}_{\bht}$.
From \Ref{acc3} we have that $\kappa_{\bht}= \frac12( \bar{\gamma}_{\bht}+  \N_{\bot}\phi )$, since it is also equal to $\bar{\gamma}_{\bht}-\frac12\bar{\d}_{\bht}$ by definition, we then  get that
\be\label{diffacc}
\bar{\delta}_{\bht} =\left(\bbs \dd \bm \bar{a}_{\bht}- \N_{\bot}\phi\right)\equiv \cL_{\bht}\hat{\eta},
\ee
is the difference between the  the screen and the Newtonian acceleration.
This difference measure to what extent the foliation and the screen are boosted with respect to each other, it therefore represent  the boost acceleration. Its expression in term of the foliation scalars is given in \Ref{dt}.

When the foliation is not orthogonal, we can express the mass density and surface tension\footnote{By definition they are related by $\rho_{\t}+2\sigma_{\t}= -\frac{\cL_{\bht}\hat\eta}{8\pi G }$} in terms of the Newtonian potential $\rho$ and the boost angle $\hat{\eta}$ as
\be
\rho_{\t}=\frac{\N_{\bot}\phi}{4\pi G}+ \frac{\cL_{\bht}\hat\eta}{8\pi G},\qquad
\sigma_{\t}=-\frac{\N_{\bot}\phi}{8\pi G}- \frac{\cL_{\bht}\hat\eta}{8\pi G},\qquad
\ee

This shows that by choosing appropriately the time dependence of the boost parameter $\hat\eta$   we can always insure  that the energy surface density or the surface tension are constant  on the screen. The choice of $\hat\eta$ essentially amount to choosing  appropriately the boost parameter $\beta$. The choice of this parameter amounts to a particular choice of foliation, it can be reabsorb into a diffeomorphism that do not move the screen. 

\section{ Dissipation}
We now want to study the evolution property of the canonical energy \Ref{E} that we have constructed.
This  energy describes a priori the energy of a open system, there is no reason therefore that this energy should be conserved in time and in general we will experience energy gain or energy loss.
In order to calculate this energy gains or losses that we call {\it dissipation} as a shorthand\footnote{As we will see the changes in energy are due to dissipation and fluxes through the screen. So here we use dissipation as a shorthand for ``energy gain or loss due to dissipation and fluxes''. We hope the reader will not get confused by this simplifying terminology.}, we proceed to express  the main conservation equation \Ref{Conseq} in a covariant form.

Let us first denote by $I_{\t}\alpha^{\mu}$  the symplectic current $\alpha^{\mu}$
evaluated on a variation $\delta g_{\mu\nu} = \cL_{\t}g_{\mu\nu}$.
From the definition (\ref{thetad}) of the symplectic current we get
\bea
I_{\t}\alpha^{\mu} &=&  [\N_{\nu},\N^{\mu}]t^{\nu} +  \N_{\nu}\left(\N^{[\nu} t^{\mu]}\right)\nn \\
&=& R_{\t}{}^{\mu} +  \N_{\nu} \kappa_{\t}^{\nu\mu}.
\eea
where we have introduced the acceleration 2-form $\kappa_{\t}^{\nu\mu} \equiv \N^{[\nu} t^{\mu]}$.
%\be
%\boxed{\kappa_{\t}^{\nu\mu} \equiv \N^{[\nu} t^{\mu]}}.
%\ee
It is interesting to write this equation  in a more suggestive manner.
Given a time flow vector $\t$ we can define the  
matter momentum current to be\footnote{ the minus sign in the matter momentum is due to the signature, the energy associated with a slice $\Sigma$ is given by
$H^{M}_{\t}=-\int_{\Sigma} T_{\t}{}^{\mu} \epsilon_{\mu} = \int_{\sigma} \sqrt{h} T_{\n\t} $} $ - T_{\t \mu}= -G_{\t\mu}/8\pi G$ and the gravitational momentum current to be:
\be
%p_{\t}{}^{\mu} \equiv - T_{\t}{}^{\mu},\qquad 
D_{\t}{}^{\mu} \equiv -I_{\t}\alpha^{\mu} + \frac{t^{\mu}}{2} R.
\ee
The canonical gravitational energy associated with a slice $\Sigma$ is simply $H^{G}_{\t} = \frac1{8\pi G} \int_{\Sigma} \sqrt{h} D_{\t\n}$. 
The conservation equation then simply reads
\be
 -D_{\t}{}^{\mu} - G_{\t}{}^{\mu}
 %=-D_{\t}{}^{\mu} - 8\pi G T_{\t}{}^{\mu}
 =  \N_{\nu}\left({\N^{[\nu} t^{\mu]}} \right).
\ee 
This expresses in particular that the total momentum form is closed, and it gives an explicit expression for its potential, in term of an acceleration tensor $\kappa_{\t}^{\nu\mu} = \N^{[\nu} t^{\mu]}$.
This conservation equation simplifies when $\t$ is a Killing field. 
In this case $\alpha_{\t}{}^{\mu}=0$ and the gravitational momentum one-form is simply $ D_{\t}{}^{\mu} = \frac{t^{\mu}}{2} R$.

It will be convenient to write this identity in the form language using the volume forms $\e,\e_{\mu},\e_{\mu\nu}$ described in the appendix.
We denote
\be
\bm{\alpha} \equiv \alpha^{\mu} \e_{\mu},\qquad \bm{\kappa}_{\t}\equiv \kappa^{\mu\nu}_{\t} \e_{\mu\nu},\qquad 
\bm{T}_{\t} \equiv T_{\t}{}^{\mu} \e_{\mu},\qquad \bm{D}_{\t} \equiv D_{\t}{}^{\mu} \e_{\mu} .
\ee
with this notation the total canonical energy  energy is given by $ H_{\t}^{G} =\frac1{8\pi G} \int_{S_{\t}} \bm{\kappa}_{\t}$ and the main conservation equation can be written
\be
\label{mainconseq}
\boxed{\rd \bm{\kappa}_{\t} = -\bm{D}_{\t} - \bm{G}_{\t}.}
\ee
The dissipation of the total canonical energy associated with a region $\Sigma$ is given by
\bea
(8\pi G) \cL_{\t} H_{\t} &=& 
\int_{\partial \Sigma_{t}} \cL_{\t}\bm{\kappa}_{\t}= 
\int_{\Sigma} \cL_{\t }\rd \bm\kappa_{\t} 
= -\int_{\Sigma} \cL_{\t} \left(\bm{D}_{\t} + \bm{G}_{\t}\right)\\
&=&-\int_{\Sigma} \rd \left(  \i_{\t} \bm{D}_{\t} + \i_{\t} \bm{G}_{\t}\right)
=- \int_{\p\Sigma}\left( \i_{\t} \bm{D}_{\t} +\i_{\t} \bm{G}_{\t}\right)\nn
\eea
where we have used that $0= \rd\bm{D}_{\t} + \rd \bm{G}_{\t}$ 
which follows from \Ref{mainconseq}.
This shows that the dissipation is purely a boundary term as expected.
Since gravity is an Hamiltonian system we do not expect any loss of energy coming from the bulk.
All the energy losses or gains comes from energy flux at the boundary two spheres.
We can evaluate the interior product  when pulled back on the surface $S$, given a one form $\alpha$ we have:
\be
\i_{\t} \alpha =  \alpha^{\mu} \i_{\t}\e_{\mu}=
\alpha^{\mu} t^{\nu} \e_{\mu\nu}  \hat{=}_{S} \sqrt{q} \alpha^{\mu} t^{\nu}(s_{\mu}n_{\nu}-s_{\nu}n_{\mu})
= -\sqrt{q}\alpha^{\mu} \ot_{\mu} = -\sqrt{q}\, \alpha_{ \bot}
\ee
where $\bot $ is  a spatial vector normal to $S$ and $\t$ and given by
\be
\ot_{\mu} = n_{\mu} (s\cdot t) - s_{\mu} ( n\cdot t),\qquad t^{\mu}\ot_{\mu}=0.
\ee
Thus the dissipation is a sum associated with the presence of each boundary spheres $\partial \Sigma = S_{o}\cup S_{i}$
\be\label{dissip1}
\boxed{\cL_{\t} H_{\t}
% = 
%\f1{8\pi G} \int_{S} \sqrt{q}\left(  G_{\t\bm{\ot}}- \alpha_{\t}^{\bm{\ot}} \right)
= -\f1{8\pi G} \int_{S_{i}}^{S_{o}} \sqrt{q}I_{\t}\alpha_{\bm{\ot}} + \int_{S_{i}}^{S_{o}} \sqrt{q}  T_{\t\bm{\ot}} .}
\ee
We have used the orthogonality of $\t$ and $\bot$ to  show that
 $R_{\t}{}^{\mu}\ot_{\mu} = G_{\t \bm{\ot}} = 8 \pi GT_{\t \bm{\ot}}$,  
 $T_{\mu\nu}$ is the energy momentum tensor. Similarly the orthogonality implies that $D_{\t\bot}=- \alpha_{\t\bot}$.
 
 In order to understand further this  conservation equation, lets suppose that  the observers are at rest with respect to the foliation, 
 i-e $\t = N \n$, then we see that $ \bm{\ot} = N \s=\bm{r}$  is going  of $\Sigma$ at the outer boundary.  
 $p_{\bm{r}}\equiv  -T_{\t \bm{r}} $ is the matter momentum density in the direction $\bm{r}$ as measured by eulerian observers. 
 We see that 
% if $p_{\bm{\ot}}$ is negative at the outer boundary, this means that momentum is entering the region $\Sigma$, then the total energy increase, while
  if $p_{\bm{\ot}}$ is positive at the outer boundary, the momentum is directed outside the region $\Sigma$ therefore the total energy decrease.
 The term  $  \int_{S} \sqrt{q}  T_{\t\bm{\ot}} =- \int_{S}\sqrt{q} p_{\bot}$ then describes the amount of energy dissipation due to matter crossing the surface $S_{t}$.
 Therefore $-\alpha_{\t \bm{\ot}}/8\pi G$ describes the  {\it gravitational } energy dissipation, it is negative  when there is energy loss due to gravitational energy leaving the system.
 
 Note that in the derivation given here, we could also evaluate the variation of $H_{\t}$ with respect to another time flow vector $\bm{\xi}$ on the screen. This is given by
  \be\label{Lxt}
%\boxed{
\cL_{\bm \xi} H_{\t}
= \f1{8\pi G} \int_{S} \sqrt{q}\left(  \cL_{\bm \xi}\kappa_{\t} +  \theta_{\bm \xi} \kappa_{\t} \right)
= -\f1{8\pi G} \int_{S} \sqrt{q}(I_{\t}\alpha_{ \bm{\xi ^{\perp}}}-{(\t \cdot\bm{\xi^{\perp}} )}R/{2})  + \int_{S} \sqrt{q}  T_{\t\bm{\xi^{\perp}}} .
%}
\ee
It is convenient to introduce a modified energy momentum tensor 
\be
\widehat{T}_{\t\bm\xi}\equiv {T}_{\t\bm\xi} - \frac{(\t \cdot\bm{\xi} )}{2} T
\ee
so the the full conservation equation reads 
\be
\boxed{\cL_{\bm \xi} H_{\t}
= \int_{S} \sqrt{q}  \left(-\frac{I_{\t}\alpha_{\bm\xi^{\perp}}}{8\pi G} + \widehat{T}_{\t\bm{\xi^{\perp}}}\right).}
\ee
It is interesting to note that for a non gravitational system, 
the previous flux equation would involve only ${T}_{\t\bm{\xi^{\perp}}}$. The contribution coming from gravity is therefore due to a ``gravitational Dissipation tensor '' $D_{\t \bxi^{\perp}}$ where
$$D_{\t \bxi}\equiv 
% {(-I_{\t}\alpha_{ \bm{\xi}}+{(\t \cdot\bm{\xi} )}R/{2})}/{8\pi G}=
-\frac{I_{\t}\alpha_{\bm\xi}}{8\pi G} - \frac{(\t \cdot\bm{\xi} )}{2} T,$$
 plays the role of a gravitational energy-momentum tensor in the conservation equation:
\be\label{dissip2}
\cL_{\bm \xi} H_{\t}
= \int_{S} \sqrt{q}  (D_{\t \bm{\xi^{\perp}}} + T_{\t\bm{\xi^{\perp}}}).
\ee
The symmetric part of this tensor is related to the dissipation of energy, since it is determined by $ D_{\t\bot}$ for a general $\t$. 
Remarkably, its antisymmetric part is related to the Poisson bracket of two  Hamiltonians. 

\subsection{Poisson bracket}
We now want to show that the antisymmetric part of the Dissipation tensor is the Poisson bracket of the Gravity Hamiltonians. 
%In order to simplify the calculation we do it here 
%only in the case of pure gravity where $T_{\a \b}$ hence $R_{\alpha \beta}$ vanish.
%In this case the dissipation tensor is simply equal to $D_{t\bxi}=-\a_{\t\bxi}$ on-shell.
We denote by $I_{\t}$ the operation of replacing a variation $\delta$ by $\cL_{\t}$, for instance
$I_{\t}\bm\alpha = \bm \alpha_{\t}$, where $\bm\alpha = \alpha^{a}\epsilon_{a}$. As a consequence we can check that 
$(\d I_{\t} +I_{\t}\d )\alpha =  \cL_{\t}\alpha + I_{\d \t} \alpha$ 
for any expression $\alpha$ which is a $n$-form on space time and a form on field space. This expression simplifies  if $\d \t =0$.
 We also denote by $\Omega $ the gravitational symplectic structure, this is a 2-form on the space of fields given by $\Omega \equiv \int_{\Sigma }\sqrt{h} \delta \alpha /8\pi G$. Lets us finally recall that $\bm{\alpha}\equiv \alpha^{\mu} \epsilon_{\mu}$ and $ \underline{\t}= \t^{\mu}\epsilon_{\mu}= \sqrt{|g|}\epsilon$. Equipped with these definition we can now compute 
\bea
(8\pi G)\d H_{\t}^{G} &= & \int_{\Sigma} \d I_{\t} \bm{\alpha} - \f12\int_{\Sigma} \delta R \underline{\t}   \nn\\
&= & - I_{\t} \int_{\Sigma} \d\bm\alpha + \int_{\Sigma} I_{\delta\t}\bm\alpha 
+  \int_{\Sigma} \cL_{\t}\bm\alpha 
- \int_{\Sigma} (\N_{\a} \alpha^{\a})\underline{\t}
 - \f12\int_{\Sigma}  (G_{\a\b} \delta g^{\a\b}) \underline{\t} - \f12\int_{\Sigma}  R \d\underline{\t}\nn\\
& = & - I_{\t} \int_{\Sigma} \d\bm\alpha + \int_{\Sigma} \left(I_{\delta\t}\bm\alpha -\frac{R}2  \underline{\d\t}\right)  - \int_{\Sigma} \sqrt{h} D_{a}(N h^{a}{}_{b} \alpha^{b})  +(8\pi G) \int_{\Sigma} \imath_{\t}\delta L^{m}
%- 4\pi G\int_{\Sigma} (G_{\a\b} \delta g^{\a \b}) \underline{\t}  -\f12\int_{\Sigma}  R t^{\mu} \d \epsilon_{\mu} 
\nn\\
&=& (8\pi G )(- I_{\t} \Omega_{\Sigma}^{G} + H^{G}_{\d \t})- \int_{ S_{i}}^{S_{o}} \sqrt{q} \alpha_{\bot}  +(8\pi G) \int_{\Sigma} \imath_{\t}\delta L^{m}
% - 4\pi G\int_{\Sigma} (G_{\a\b} \delta g^{\a \b}) \underline{\t} -\f12\int_{\Sigma}  R t^{\mu} \d \epsilon_{\mu} 
\eea
where $ \Omega_{\Sigma}^{G}$ denotes the gravitational part of the symplectic form, $\Delta S \equiv S_{o}-S_{i}$ and we have used that $\delta L^{m}=-\frac12\sqrt{g} T_{\alpha\b}\delta g^{\a\b}$. 
In the second line we have used that $\d( \sqrt{g} R) =  \sqrt{g} G_{ab}\d g^{ab} + 2  \sqrt{g}\N_{a}\alpha^{a}$.
We also used in the third line that
$\sqrt{g} \N_{\mu} \alpha^{\mu} = \cL_{N\n}\bm \alpha +\sqrt{h} D_{a}(Nh^{a}{}_{b}\alpha^{b})$.
This implies that given a form $\bm{\omega} = \omega^{\a}\epsilon_{\a}$ and denoting $\omega_{\n}=\omega^{\a}n_{\a}$ we have that
\be
\int_{\Sigma} (\N_{\a}\omega^{\a}) \underline{\t} = \int_{\Sigma}\cL_{\t} \bm \omega +\int_{\partial \Sigma} \omega_{\bot}\sqrt{q}.
\ee
%THus in summary we have for the gravitational variation that 
%\be
%\d H_{\t}^{G} - H^{G}_{\d \t} = - I_{\t} \Omega_{\Sigma}^{G} - \int_{ S_{i}}^{S_{o}} \sqrt{q} \left(\frac{\alpha_{\bot}}{8\pi G}\right)
%+ \int_{\Sigma} \imath_{\t}\delta L^{m}.
%\ee
%where we used that $ \imath_{\t}\bm{\alpha}  = - \sqrt{q} \alpha^{\mu}\ot_{\mu}$.
Let us assume for simplicity that we are in the context of pure gravity, so that $L^{m}=0$ and $H^{G}_{\t}=H_{\t}$.
This shows in this context that 
%We also consider the variation of the matter Lagrangian which gives similarly
%\be
%\d H_{\t}^{m}-H^{m}_{\d \t} = - I_{\t} \Omega_{\Sigma}^{m} - \int_{\Sigma} \imath_{\t}\delta L^{m}.
%\ee
%Adding the two contribution together we get for 
the total Hamiltonian  variation is 
\be
\boxed{( H_{\d \t}-\d H_{\t})
=  I_{\t} \Omega_{\Sigma}  +\frac1{8\pi G} \int_{ S_{i}}^{S_{o}} \sqrt{q}{\alpha}_{\bot}.} \label{dH}
\ee
Contracting this with $I_{\bxi}$ and using the definition  $I_{\bxi}\delta H =\cL_{\bxi}H$,
and the definition of the Poisson bracket $ I_{\bxi}I_{\t} \Omega_{\Sigma} =\{H_{\t},H_{\bxi} \}$,
%together with  $H_{\t} = H^{G}_{\t} - \int_{\Sigma }\bm{T}_{\t}$ 
we get 
\be
%\boxed
H_{[\t,\bxi]}- { \cL_{\bxi} H_{\t}
% &=& -I_{\bxi}I_{\t} \Omega_{\Sigma} +H^{G}_{[\bxi,\t]} 
%-  \int_{\Delta S}  \f{\alpha_{\bxi \bot}}{8\pi G} \sqrt{q}
% + \int_{\Sigma}  \N^{\a}(T_{\a\b} \xi^{\b} )\sqrt{g}- \int_{\Sigma }\cL_{\bxi}\bm{T}_{\t}
% - \f12\int_{\Sigma}  R(\N_{\a}\xi^{\a}) \underline{\t}
%  \nn\\
% &=&  -\{H_{\t}^{G},H_{\bxi}^{G} \}+ \frac1{8\pi G} \int_{\Sigma} \bm\alpha_{[\bxi,\t]}  - \f1{8\pi G}\int_{\Delta S} \sqrt{q} {\alpha_{\bxi \bot} }
% +  \int_{\Delta S} \sqrt{q} T_{\bxi\bot}  
% +  \int_{\Sigma} (\cL_{\t} \bm{T}_{\bxi} - \cL_{\bxi}\bm{T}_{\t})\nn\\
 = \{H_{\t},H_{\bxi} \} + \frac1{8\pi G} \int_{\Delta S}\sqrt{q} {I_{\bm\xi}\alpha_{ \bot} }.
%  +  \int_{\Sigma} (\cL_{\t} \bm{T}_{\bxi} - \cL_{\bxi}\bm{T}_{\t} - \bm T_{[\t,\bxi]}) 
  }\label{dH1}
\ee
Taking the difference between this and \Ref{Lxt} gives 
\be\boxed{
\{H_{\t},H_{\bxi} \}= H_{[\t,\bxi]} +\f1{8\pi G}  \int_{\Delta S}\sqrt{q}\left( I_{\t}\a_{\bxi^{\bm\perp}}- I_{\bm\xi}\alpha_{ \bot} \right).
}\label{Poisson}
\ee
Thus we see that Poisson bracket algebra, is now {\it anomalous} due to the presence of the boundary and that the combination $(I_{\t}\a_{\bxi^{\bm\perp}}- I_{\bm\xi}\alpha_{ \bot})/8\pi G = -D_{\t\bxi^{\bm\perp}}+D_{\bxi \bot}$ gives the anomaly,  in the context of pure gravity.

\section{The thermodynamical balance equation}
From the dissipation equation \Ref{dissip1} and the on shell evaluation $H_{t}=\int_{S}\sqrt{q}\kappa_{\t} $ we have that:
\be\label{main1}
 \cL_{\t}\left(\sqrt{q} \kappa_{\t}\right) +  I_{\t}\alpha_{\bot}  {=}   8\pi G \sqrt{q} T_{\t \bot}
\ee 
%In practice $\d=\cL_{\xi}$ where $\xi$ is a foliation preserving vector field.
The LHS is evaluated in the appendix for a vector field $\t=\bht + \bphi$ which is foliation preserving.
Here $\bht =\rho \bbn$ and $\bphi$ is tangent to $S$.
By separating the contribution that comes from $ \bht$ from the one that comes from $\bphi$ we get two equations \Ref{et}, \Ref{mt}. The first one is an energy balance equation, the second one is a momentum balance equation. As we will see they have a  natural thermodynamical interpretation 
and leads to what is  a generalized first law and a generalized Cauchy momentum equation.
 We first look at the energy balance.

\subsection{The energy balance}\label{ebal}

The energy balance equation  is a consequence of \Ref{et} derived in the appendix.
This equation reads:
\be\boxed{
 (\cL_{\bht}  + \theta_{\bht}) \theta_{\bot}- \bar{\gamma}_{\bht} \theta_{\bht}
   + \bm{\widehat\Theta}_{\bot}:\bm\Theta_{\bht} + (8\pi G)T_{\bht\bot}   -\theta_{\bot}\cL_{\bht}\phi 
   - \frac12\bm\rd \dd (\bm{q}[\bht,\bot])  = 0.
   %- \rd_{a}(\rho \hat\d^{a}_{\bbs})
 }\label{et}
 \ee
 Here $(\bm\Theta_{\bht},\bm{\Theta}_{\bot})$ are the extrinsic deformation tensors,
 $(\theta_{\bht},\theta_{\bot})$ denotes their trace, $\phi =\ln \rho$ is the newtonian potential,
 while $\bar{\gamma}_{\bht}$ is the radial acceleration. The hat on $\bm\Theta_{\bm\ell}$ means 
 $\widehat{\Theta}^{AB}_{\bm\ell}\equiv {\Theta}^{AB}_{\bm\ell}- q^{AB} \theta_{\bm\ell}$.
 We can write this balance equation in terms of the thermodynamical quantities that we have introduced 
 First let us recall that the density of internal energy and the surface tension are  given by 
 \be
 \epsilon \equiv -\frac{\theta_{\bot}}{8\pi G},\qquad \sigma_{\t}\equiv -\frac{\bar{\gamma}_{\bht}}{8\pi G}.
 % -\frac{\cL_{\bht}\bar{\eta}}{16\pi G}
 \ee
 The previous equation can therefore be written as an energy conservation:
 \be\boxed{
 (\cL_{\bht}  + \theta_{\bht}) \epsilon =  \sigma_{\t} \theta_{\bht}
   + \frac{\bm{\widehat\Theta}_{\bot}}{8\pi G}:\bm\Theta_{\bht} + T_{\bht\bot}   +\epsilon\cL_{\bht}\phi 
   - \frac12\bm\rd \dd \left(\frac{\bm{q}\cdot[\bht,\bot]}{16\pi G}\right)  = 0.
   %- \rd_{a}(\rho \hat\d^{a}_{\bbs})
 }\label{et}
 \ee
Each terms appearing in this equation have a natural thermodynamical interpretation.
In order to see this lets integrate the previous equation over $S$ and
 let us introduce several quantities:
 First we define 
 \be
 \dot{E}_{\mathrm{M}} \equiv \int_{S}\sqrt{q} T_{\bht\bot}.
 \ee
 This is the rate of matter energy flowing through the screen. It is positive if matter is leaving the external region and entering the screen. 
 This corresponds to a work term due to the transport of matter.
 
 Lets also define the rate of gravitational dissipation due to gravitational radiation.
 It is given by
 \be
 \dot{\cal Q} \equiv\frac1{8\pi G} \int_{S}\sqrt{q} ( \bm \Theta_{\bht} :\bm \Theta_{\bot} - \theta_{\bht} \theta_{\bot})
 =\int_{S}\sqrt{q}  \bm \tau :\bm \Theta_{\bht} .
 \ee
 This correspond to an internal  heat production term due to gravitational dissipation.
 It is analogous the the entropy production term $T\dot{S}$ in a fluid where $ \Theta_{\bht} $ plays the role of the rate of strain tensor and $$\bm{\tau}\equiv  \frac1{8\pi G}\bm{ \widehat{\Theta}}_{\bot}$$ plays the role of the viscous stress tensor \cite{degroot,YukiL1}.

 The {\it internal energy} is defined, as we have seen, to be proportional to the radial expansion:
 \be
 U\equiv \int_{S} \sqrt{q} \epsilon, \mrm{with}  \epsilon = - \frac{\theta_{\bot}}{8\pi G}. 
 \ee
 We also denote element of area and its  rate of change  by
 \be
 \rd A =\sqrt{q} \rd^{2}x,\quad  (\cL_{\bht} \rd A) =  \theta_{\bht} \rd A 
 \ee
 The term $\sigma_{\t} (\cL_{\t} \rd A)$ is a work term due to the presence of surface tension.
 
 The final term we want to analyze corresponds to the rate of change of the Newtonian energy of the screen.
 We define the newtonian gravitational energy of the screen to be 
 \be
 E_{\mathrm{N}} \equiv \int_{S}\sqrt{q} \epsilon \phi.
 \ee
 This means that  $\epsilon$, which is the internal energy density, also represents the inertial mass density.
Therefore we see that the change of the newtonian energy due to the time variation of the potential is given by
 \be
 \dot{E}_{\mathrm{N}} = \int_{S}\sqrt{q} \epsilon \cL_{\bht} \phi 
 %=  
 %\int_{S}\sqrt{q} \left(  - \frac{\theta_{\bot}}{8\pi G} \cL_{\bht} \phi + \frac{\Theta_{\bht}}{8\pi G}   \cL_{\bot}\phi\right).
 \ee
 which is the term entering the balance equation.
 
 We can therefore write the integrated balance equation as 
 \be\boxed{
 \cL_{\bht} U =  \dot{\cal Q}  + \int_{S}\sigma_{\t}\cL_{\bht}\rd{A} + \dot{E}_{\mathrm{M}} + \dot{E}_{\mathrm{N}} 
 }\label{first law}
 \ee
 This is the generalized first law of thermodynamics for a general screen.
 By integrating it out over a small amount of time and assuming for simplicity that $\sigma_{\t}$ is constant we get
 \be
\rd U =   \delta{\cal Q}  + \sigma_{\t} \rd{A} + \delta {E}_{\mathrm{M}} + \delta {E}_{\mathrm{N}} 
 \ee
 here we differentiate between the total differential $\rd$ of potential and the infinitesimal variation $\delta$ of a quantity that do not represent the state of the system.
 This formula justifies a posteriori the identification of $U$ as the internal energy.
 It expresses the variation of the internal energy $U$ in terms 
 of the rate of work done on the system plus the rate of heat production, that is $\delta U = \delta W + \delta \cal{Q}$. 
Here the work terms are three-fold, first there is a Newtonian work term due to the fact that the internal energy is also the inertial mass for the newtonian potential.
There is also a matter work term due to matter flowing in or out of the system and there is finally a 
work term due to the change in size of the system. Since the system is two dimensional and posses surface tension  this work term 
reads $\sigma \rd A $. When $\sigma $ is negative (for outer screens for instance) one should interpret it as a 2 dimensional pressure $p_{2d}=-\sigma$ and the work term is the 2 dimensional analog of  $-p \rd V$.
In this case the surface tension acts as a 2d pressure term from the point of view of the screen. 

Note that if one looks at transformations that do not produce any heat (i-e $\delta {\cal{Q}}=0$ hence not gravity wave production) and transformations that do not change the internal energy $\rd U =0$ and the Newtonian energy; the previous relation can be written as  
$$\rd E_{\mathrm{M}}= p_{2d} \rd A.$$
This is this relation that has been interpreted in the literature as a Clausius relation leading to the identification $T\rd S =  p_{2d} \rd A$. We see that this interpretation is valid only in a very restricted context.
In general the entropy production term for the 2-dimensional thermodynamical system  is proportional to $\delta \cal{Q}$ hence measure the production of gravity wave. The presence of a term $\sigma \rd A$ is interpreted as a work term due to the presence of a 
surface tension or 2dimensional pressure
%\footnote{ In usual thermodynamical systems with interface 
%the surface tension $\sigma$ plays the same role as a negative 2dimension pressure term.}
 
\subsection{The momentum balance}
In the appendix we evaluate the  local balance equation for the momentum, we get:
%Taking the difference between \Ref{Dej} and \Ref{Dj},
%and using Einstein equation we get the local balance equation for the Angular momentum
%\be\label{Jeq}
%\cL_{\bht}j_{\bm \varphi} + \Theta_{\bht} j_{\bm \varphi} 
%=\bm \varphi \cdot  \left( 
%  - \bm \rd \kappa_{\bht}  
%+ \bm \rd \cdot \bm \Theta^{\bot} - \bm \rd  \Theta^{\bot} \right) -  8 \pi G T_{\bm\varphi \bot}
%+(\Theta_{\bot} {\rd_{\bm \varphi}\phi} + \Theta_{\bht} \bar{\delta}_{\bm \varphi})
%   + \rd^{A} \left(\frac12 \cL_{\bot} \varphi_{A} -\varphi^{A}\kappa_{\t}\right)
%\ee
\be
\boxed{ \varphi^{a}(\cL_{\bht}+\theta_{\bht}) \bm{\bar\pi}_{a}=
 \varphi^{a}\left( \left( -\brd \bar{\gamma}_{\bht} + \brd\cdot \bm{\widehat{\Theta}}_{\bot}\right)_{a}
   +\theta_{\bot}\rd_{a}\phi - (8\pi G) T_{a \bot} \right)
       - (8\pi G)\rd_{a}v_{\bphi}^{a }
}\label{mt}
\ee
where $ (8\pi G) v_{\bphi}^{a }\equiv \varphi_{a} (\theta_{\bot} - \kappa_{\t})   +\frac12 q^{ab}( \cL_{\bot} \varphi_{b})$. $\phi=\ln \rho$ is the Newtonian potential ${\bar\pi}_{A}=\bbs\dd\N_{A}\bbn$ is the normal one-form in the screen frame, $\bm{\widehat{\Theta}}_{\bot}= \bm{{\Theta}}_{\bot}- \bm q \theta_{\bot}$ and $\theta_{\bot}$ is the trace of the extrinsic tensor $\bm{{\Theta}}_{\bot}$ .
We can write this equation in terms of the thermodynamical quantities like the
 internal energy density $\epsilon$,
 %= -\frac{\theta_{\bot}}{8\pi G}$, 
 the 2d surface tension $\sigma_{\t}$, 
 The viscous stress tensor $\bm{\tau}$
 %= \frac{\bm{\widehat\Theta_{\bot}}}{8\pi G}$, 
 and  the momenta density
$
\bm{p},
$
where 
\be
\epsilon= -\frac{\theta_{\bot}}{8\pi G},\qquad
\sigma_{\t}= -\frac{\bar\gamma_{\bot}}{8\pi G},\qquad 
\bm{\tau}= \frac{\bm{\widehat\Theta_{\bot}}}{8\pi G},\qquad
\bm{p} = \frac{\bm{\bar\pi}}{8\pi G}.
\ee
The momentum conservation equation reads 
\be
\boxed{ \varphi^{a}(\cL_{\bht}+\theta_{\bht}) p_{a}=
 \varphi^{a}\left[ \left( \brd \sigma_{\t}+ \brd \cdot \bm\tau \right)_{a}
   +\epsilon \rd_{a}\phi - T_{a \bot} \right]
       - \rd_{a}v_{\bphi}^{a }.
}\label{mt2}
\ee
We now look at the integrated version of this balance equation.

First, we define the total  momenta and its time variation as 
\be
 \bm{P}_{\bphi}\equiv  \int_{S} \sqrt{q} \varphi^{A} p_{A},\qquad 
 \dot{\bm{P}}_{\bphi}\equiv  \int_{S} \sqrt{q} \varphi^{A} (\cL_{\bht}+\theta_{\bht})p_{A}.
\ee
We also define the force acting on the screen  due to the flow of matter across the horizon
\be
{F}_{\bphi}^{\mathrm{M}} \equiv -\int_{S} \sqrt{q} T_{\bphi \bot}
\ee
Indeed, Lets integrate the conservation of the energy momentum tensor  over a space-time region 
R bounded by the screen and an initial and final slice.
The matter momenta on each slice is given by $p_{A} = - \int_{\Sigma} \sqrt{h} T_{\n A} $.
thanks to the conservation equation
$\N_{\mu} T^{\mu\nu}=0$, and the Gauss law, we can evaluate the variation of momenta to be
\be
\Delta p_{A} =- \int_{\bar{\Sigma}} \sqrt{\bar{h}} T_{\bbs A} = -\int \rd T \left(\int_{S}\sqrt{q} T_{\bot A}\right). 
\ee
where $\bar{h}$ is the induced metric on the screen. 
Therefore the rate of change of momenta in the direction $A$ per unit time, which gives the force acting on the system is given by 
$F_{A}^{\mathrm{M}}= - \int_{S} \sqrt{q} T_{\bbs A}$.

Next, we define the Newtonian force acting on the system to maintain the screen in place, it is given by
\be\label{Fnewt}
F^{\mathrm{N}}_{A} \equiv -\int_{S} \sqrt{q}  \epsilon \rd_{A} \phi.
\ee
Here $\epsilon = -\frac{\Theta_{\bot}}{8 \pi G} $ is the internal energy density. It is also as we have see the screen inertial mass density. Thus  $-\epsilon \rd_{A} \phi   $ is simply the usual newtonian force
acting on a system of inertial mass density  $\epsilon$. 

Finally, the first term in \Ref{mt2}  is a 2d pressure force term $\rd \sigma = -\rd_{a} p_{2d}$.
It is the force term due within the presence of surface tension. It forces the fluid to flow in the direction of high surface tension. It is responsible for Marangoni flows.
The second term 
is identical to a term of viscous force due to a stress tensor ${\tau}^{AB}= (\bm\Theta_{\bot} -\Theta_{\bot} \bm{q})^{AB}/(8\pi G)$ confirming again this interpretation.
These are the  convection terms.

 Integrating this equation on $S$ and assuming that the vector $\varphi^{A} $ is conserved in time we get 
\be
 \dot{ P}_{\bphi} =- \int_{S} \varphi^{A} \rd_{A} \sigma_{t} +  \int_{S} \varphi_{A} \rd_{B} \tau^{AB} + {F}_{\bm \varphi}^{\mathrm{M}}+ F^{\mathrm{N}}_{\bm \varphi} 
\ee
This equation is identical to the conservation of momenta in a general non-equilibrium thermodynamical system.
It confirms in particular the interpretation of $\sigma_{\t}$ as a surface tension and $\bm\tau$ as a viscous stress tensor.

\section{Conclusion}
  In this work we have studied in great detail the definition of energy for a gravitational system
  in the context of a 2+2 decomposition of spacetime.
  We have seen that a general gravitational screen possesses gravitational analogs of 
  a surface tension, an internal energy and a viscous stress tensor.
  These data enters the conservation of energy and momenta described from the point of view of the screen.
  And these shows that the gravitational equations projects themselves on the screen as non-equilibrium conservation equation for the screen degrees of freedom.
  This analysis provides a first sets of clues toward understanding gravitational systems as thermodynamical systems.
  The analogy presented here is purely classical and needs to be deepened \cite{YukiL1}.
  One of the fundamental puzzle to elucidate concerns understanding the nature of the surface tension of gravitational screens.
  
  The idea that gravity and thermodynamics are related subject is not new.
A thermodynamical interpretation of local null screens has been developed by Jacobson et al \cite{Jacobson} and by 
Padmanabhan \cite{Padmanabhan}. Also Verlinde \cite{Verlinde} has proposed to derive gravity from 
entropic arguments and equipartion of energy associated with  holographic screens.
In these works however, it is {\it postulated} that the surface tension is a local temperature and hence it is {\it assumed} that the work term $\sigma \rd A$  is due to an entropy variation.
Whereas this seems to be well established for bifurcated killing horizons \cite{ Bardeen, WaldE}, 
it is higly speculative  at this point to generalize this interpretation beyond that very particular situation.
Our analysis clearly shows  that we can associate a surface tension to a general gravitational screen and that entropy production is due to gravity waves (see also \cite{Eling,Chirco,Yokokura,Shimada} for similar consideration on gravity wave and null screens). What does it tells us about the entropy and temperatures of a general gravitational system is still an open problem.
One direction of investigation that we intend to pursue in the future is to deepen the possible 
relationship between entanglement energy and the boost energy introduced here.

One should also mention that it will be also interesting to understand the relationship of our general screen approach with the gravity-fluid correspondence \cite{Hubeny} developed in AdS/CFT.
Although there the non relativistic equations appears  from a non relativistic expansion of solutions of general relativity which is a very different setting.

Finally the idea that non-equilibrium thermodynamics might be a key toward a theory of quantum gravity constituents via a fluctuation-dissipation theorem, is a fascinating idea that has received little attention yet
\cite{HawkingH, Sciama,Pranzetti}. I hope that such an investigation  could open a new avenue towards quantum gravity.

\section*{Acknowledgement}
I am deeply grateful to E. Bianchi for many insight-full exchanges and feedback, to D. Minic for many encouragements and feedback on the manuscript and to Y. Yokokura for endless discussions about this subject and detailed feedback on the manuscript.
I have benefited the insight, feedback and encouragements of many colleagues. Many thanks to 
H. Haggard, J. Kowalski-Glikman, L. Lehner, L. Smolin,  for their input.
This work started while I was a visitor at Yukawa institute in Kyoto, I thanks deeply N. Sasakura for an inspiring stay in this magnificent city.

This research was supported in part by Perimeter Institute for Theoretical Physics. Research at Perimeter Institute is supported by the Government of Canada through Industry Canada and by the Province of Ontario through the Ministry of Research and Innovation. This research was also partly supported by grants from NSERC.   
  \appendix
     
\section{ Dissipation: The derivation}
From \Ref{dH1} and the on shell evaluation $H_{t}=\int_{S}\sqrt{q}\kappa_{\t} $ we have that:
\be\label{main1}
\frac{1}{8\pi G}\int_{S}\left( \cL_{\t}\left(\sqrt{q} \kappa_{\t}\right) +  I_{\t}\alpha_{\bot} \right) \hat{=}   \int_{S}\sqrt{q} T_{\t \bot}
\ee 
where $\hat{=}$ means that we evaluate the expression on-shell. 
%In practice $\d=\cL_{\xi}$ where $\xi$ is a foliation preserving vector field.

In this section we want to evaluate  LHS of this expression. The main task comes from the evaluation of  the dissipation tensor $ I_{\t }\alpha_{\bot}.$ We start  from the general expression \Ref{theta1} for the contraction of the symplectic potential in the normal direction $\bot =\rho \bbs$. If one specialise this formula to variations $\delta$ that preserves the foliation (i-e $\bar{h}^{ab}\delta \bs_{b}=0$)  we have that 
\be
-\sqrt{|\bar{h}|} \alpha_{\bbs}= \delta\left(\sqrt{|\bar{h}|} {K}_{\bbs} \right) + \frac{\sqrt{|\bar{h}|}}{2}\left(K_{\bbs}^{ab}- \bar{h}^{ab} K_{\bbs} \right) \delta \bar{h}_{ab} -\sqrt{|\bar{h}|} \bar{D}_{a} {\delta}_{\bbs}^{a}
\ee
 where $ \bar{h}_{ab}= q_{ab} - \bn_{a}\bn_{b}$ is the metric on the timelike screen, 
 $K_{\bbs}^{ab} = \bar{h}^{a}{}_{a'}\bar{h}^{b}{}_{b'}\nabla^{a'} \bs^{b'}$ is the extrinsic tensor for the spacelike normal $\bbs$, and $\delta_{\bbs}^{a}\equiv \frac12(\bar{h}^{a}{}_{b}\delta \bs^{b} + \bar{h}^{ab}\delta \bs_{b})= \frac12 \bar{h}^{a}{}_{b}\delta \bs^{b}$ is the projected variation.
 In practice $\d=\cL_{\t}$ where $\t= \bht + \bphi$,  is a foliation preserving vector field, that is $\bht =\rho \bbn$ and $\bphi$ is a vector tangent to $S$.
 
 We decompose the extrinsic tensor in its space and time components
 \be
 K_{\bbs}^{ab} =  -\bar{\gamma}_{\bbn}\bn^{a}\bn^{b} +\bpi^{a}\bn^{b} + \bpi^{b}\bn^{a} + {\Theta}^{ab}_{\bbs},
 \ee
 where $\bar{\gamma}_{\bbn} = \bbs\dd\N_{\bbn} \bbn$ is the radial acceleration, $\bpi_{a}=q_{a}{}^{b}\bbs\cdot \N_{b}\bbn$ the normal connection and ${\Theta}^{ab}_{\bbs}$ the $\bbs$ extrinsic curvature tensor.
 This implies that $ K_{\bbs} =  \bar{\gamma}_{\bbn} + {\theta}_{\bbs}$, and
 \bea
 %K_{\bbs} &=&  \bar{\gamma}_{\bbn} + {\theta}_{\bbs},\\
 \left(K_{\bbs}^{ab}- \bar{h}^{ab} K_{\bbs} \right) = \theta_{\s}n^{a}n^{b} +\bpi^{a}n^{b} + \bpi^{b}n^{a} + \widehat{\Theta}^{ab}_{\bbs}-q^{ab}\bar{\gamma}_{\bbn}.
 \eea  
where $\widehat{\Theta}^{ab}_{\bbs} \equiv {\Theta}^{ab}_{\bbs}- q^{ab } \theta_{\bbs}$.
% Lets compute $\bar{\Pi}^{ab} \delta \bar{h}_{ab}$ where 
% $n_{a}n_{b}\Pi^{ab}= \theta_{\s}$ and  $ n_{b}\Pi^{ab} = n_{b} \N^{a} s^{b}= -\bpi^{a}$
% $$\Pi^{ab} = \theta_{\s}n^{a}n^{b} +\bpi^{a}n^{b} + \bpi^{b}n^{a} + \widehat{\Theta}^{ab}$$.
 Then we consider variations that preserves the screen foliation (i-e $ q^{ab} \delta \bn_{b}=0$) 
for which  we have\footnote{ The expression for $\bn^{a}\delta \bn_{a}$ follows from instance from the 
 expression for the normal form and vector: $\bn_{a}\rd x^{a}= -\rho \rd T + \tau\sinh\eta \rd R$ and $\rho \bn^{a}\partial_{a}= \partial_{T} - \varphi^{A}\partial_{A}$ and then use that for a foliation preserving variation $\d \rd T =0$.}  
 \be
  \bn^{a}\bn^{b}\delta \bar{h}_{ab}= 2\bn^{a}\delta \bn_{a}= -2\frac{\delta \rho}{\rho} =-2\d \phi,\qquad
 \bn^{a}\bpi^{b}\delta \bar{h}_{ab}= - \bpi_{a} \d \bn^{a}.
 \ee
 Here we have introduce the Newtonian potential $ \phi \equiv \ln \rho$.
  Therefore we get that 
 \bea
 \left(K_{\bbs}^{ab}- \bar{h}^{ab} K_{\bbs} \right) \delta \bar{h}_{ab} =-2 \bar{\gamma}_{\bbn} \theta
 + \widehat{\Theta}^{ab}_{\bbs} \delta q_{ab} 
 -2\theta_{\bbs}\delta\phi - 2 \bpi_{a}\d \bn^{a}  
   \eea
 where we have introduced the notation $q^{ab}\delta q_{ab}\equiv 2\theta$.
 The other term to consider is 
 \be
\rho \bar{D}_{a}\delta_{\bbs}^{a}
 =\rho \bar{D}_{a}(\bn^{a} \bar{\delta}) + \rd_{a}(\rho \hat\d^{a}_{\bbs})
 = (\cL_{\bht} + \theta_{\bht})\bar{\delta} + \rd_{a} \hat\d^{a}_{\bot} 
 %+ \bm\rd\dd \bm{\hat\d}_{\bot},
 \ee
 where $\bar{\delta}\equiv\frac12( \bs_{a}  \delta \bn^{a}+ \bs^{a}\delta \bn_{a})$ is the normal component and 
 $\hat\d^{a}_{\bot}= q^{a}{}_{b}\delta_{\bot}^{b}= \rho q^{a}{}_{b}\delta_{\bbs}^{b}$ the tangential component.
 The last term entering the dissipation tensor is
 \be
  \delta\left(\sqrt{|\bar{h}|} \bar{K}_{\bbs}\right) 
  %= \d(\bar\gamma_{\bbn} + \theta_{\bbs}) + (\bar\gamma_{\bbn} + \theta_{\bbs}) (\delta \varphi + \f12 q^{ab}\delta q_{ab}),
  = \d\left(\sqrt{q} \rho (\bar\gamma_{\bbn} + \theta_{\bbs})\right).
 \ee
 Finally, we have to take into account the Hamiltonian variation.
 For $\t =\bht +\bphi$ we have $\kappa_{\t} = \rho (\bar{\gamma}_{\bbn}-\d_{\bbn}) - \bphi\dd\bj$, where 
 $\bar{\gamma}_{\bht}= \bbs\dd\N_{\bht}\bbn$ is the radial acceleration and $\bar\d_{\bbn}=I_{\bbn}\bar{\delta}$.
 we can evaluate the LHS of \Ref{main1}, that is $\d\left(\sqrt{q} \kappa_{\t}\right) +  \alpha_{\bot}$ to be equal to:
 \bea
 %\d\left(\sqrt{q} \kappa_{\t}\right) +  \alpha_{\bot}
 & &  \d\left(\sqrt{q}  (\kappa_{\t} - \rho K_{\bbs})\right) -
   \frac{\sqrt{|\bar{h}|}}{2}\left(K_{\bbs}^{ab}- \bar{h}^{ab} K_{\bbs} \right) \delta \bar{h}_{ab} 
   +\sqrt{|\bar{h}|} \bar{D}_{a} {\delta}_{\bbs}^{a} \nn\\
  &=& -\d\left(\sqrt{q} [\rho(\theta_{\bbs}+\d_{\bbn})+ \bphi\dd\bj]\right)- \sqrt{q}\rho
  \left(
  -\theta_{\bbs}\delta\phi -  \bpi_{a}\d \bn^{a}  + \frac12 \left(\widehat{\Theta}^{ab}_{\bbs} -q^{ab}\bar{\gamma}_{\bbn}\right) \delta q_{ab} -
   (\cL_{\bht} + \theta_{\bht})\bar{\delta} \right) +\sqrt{q}  \rd_{a}(\rho \hat\d^{a}_{\bbs})\nn.\\
   &=& -\sqrt{q}\left((\d+\theta) (\theta_{\bot}+\bphi\dd\bj + \bar\d_{\bht})  -   (\cL_{\bht} + \theta_{\bht})\bar{\delta}
   - \bar{\gamma}_{\bht} \theta
   + \frac12 \widehat{\Theta}^{ab}_{\bot}\delta q_{ab}
   -\theta_{\bot}\delta\phi -  \bpi_{a}\d \hat{t}^{a} 
   - \rd_{a}( \hat\d^{a}_{\bot}) \right)\label{maind}
   %\\
%  &\hat{=} & \sqrt{q}2\theta_{\bbs}\delta\varphi - 2 \bpi_{a}\d \bn^{a}  + \widehat{\Theta}^{ab} \delta q_{ab} +
%   \bar{D}_{a}(\bn^{a} \bar{\delta}) + \rd_{a} \hat\d\bs^{a}
 \eea
 %We used that $ \rho \bar{D}_{a}(q^{a}{}_{b} \d^{b}) = \rd_{a}(\rho q^{a}{}_{b}\d^{b})$.
 We first specialize to the case where  $\bphi=0$ hence $\t=\bht$ and then contract the previous expression with $I_{\bht}$ so that $ I_{\bht}\d\alpha=\cL_{\bht}\alpha$ is a time diffeomorphism.
 The previous expression is on-shell equal to $(8\pi G)\sqrt{q}T_{\bht\bot}$ and therefore the energy balance equation simplifies to 
 \be\boxed{
 (\cL_{\bht}  + \theta_{\bht}) \theta_{\bot}- \bar{\gamma}_{\bht} \theta_{\bht}
   + \bm{\widehat\Theta}_{\bot}:\bm\Theta_{\bht} + (8\pi G)T_{\bht\bot}   -\theta_{\bot}\cL_{\bht}\phi 
   - \bm\rd \dd (\rho^{2}\bm j)  = 0.
   %- \rd_{a}(\rho \hat\d^{a}_{\bbs})
 }\label{et}
 \ee
We have used that $I_{\bht}\hat\d^{a}_{\bot} = \frac12 q^{a}{}_{b}[\bht,\bot]^{b} = \rho^{2} j^{a}$, with $\bm{j}$ the twist vector.
\subsection{Momentum balance derivation}\label{mbal}
We now contract \Ref{maind}  with $I_{\bht + \bphi}$ and subtract from it the contraction with $I_{\bht}$ which gave the previous equation. This difference is  equal on-shell to $(8\pi G)\sqrt{q}T_{\bht\bphi}$,  and we get 
\be
(\cL_{\bht}+\theta_{\bht}) (\bphi\cdot \bj-\bar{\delta}_{\bphi}) +(\cL_{\bphi}+\brd\cdot \bphi) (\theta_{\bot}+ \bphi\cdot \bj+\bar\d_{\bht}) 
- \bar{\gamma}_{\bht} \brd\cdot \bphi
+  \widehat{\Theta}^{ab}_{\bot}\N_{a}\varphi_{b} 
   -\theta_{\bot}\cL_{\bphi}\phi -  \bm\bpi\cdot [\bphi, \bht] 
   - \frac12\bm\rd\cdot [\bphi,\bot] + T_{\bphi \bot}=0 \nn
\ee
where we used that $I_{\bphi}\theta = \frac12 q^{ab}\cL_{\bphi}q_{ab} = q^{ab}\N_{a} \varphi_{b}=\rd_{a}\varphi^{a}$, and that $\bar{\delta}_{\bphi}=I_{\bphi}\bar{\delta}$.
It is possible to simplify greatly this expression: First one integrate by part $-\bar{\gamma}_{\bht} \brd\cdot \bphi = -\brd ( \bphi \bar{\gamma}_{\bht}) + \bphi\cdot \brd \bar{\gamma}_{\bht}  $ and we use that 
$(\cL_{\bphi}+\brd\cdot \bphi) \alpha = \brd ( \bphi  \alpha)$ is a total derivative.
This gives 
\be
(\cL_{\bht}+\theta_{\bht}) (\bphi\cdot \bj-\bar{\delta}_{\bphi} ) 
+ \brd \left( \bphi (\theta_{\bot} + \bphi\cdot \bj + \bar\d_{\bht}- \bar{\gamma}_{\bht}  ) \right)
+ \bphi\cdot \brd \bar{\gamma}_{\bht} 
   +  \widehat{\Theta}^{ab}_{\bot}\N_{a}\varphi_{b} 
   -\theta_{\bot}\cL_{\bphi}\phi -  \bm\bpi\cdot [\bphi, \bht] 
    - \frac12\bm\rd\cdot [\bphi,\bot] + T_{\bphi \bot}=0 \nn
\ee
One also integrate by part  $  \widehat{\Theta}^{ab}_{\bot}\N_{a}\varphi_{b}  =  \rd_{a}(\widehat{\Theta}^{ab}_{\bot}\varphi_{b} ) -  (\rd_{a}\widehat{\Theta}^{ab}_{\bot})\varphi_{b} $ and then use that 
$$\widehat{\Theta}^{ab}_{\bot}\varphi_{b} - [\bphi,\bot]^{a} = \frac12\left( \N_{\bphi}\bot + \bphi\cdot\bm{\N}\bot - \N_{\bphi}\bot + \N_{\bot}\bphi \right)^{a}= \frac12 q^{ab} \cL_{\bot} \varphi_{b}.$$
One also use that 
$\bphi\cdot \bj-\bar{\delta}_{\bphi}  = -\bm\bpi \cdot \bphi$ and that 
$$- \cL_{\bht}\bm\bpi \cdot \bphi -\bm\bpi\cdot [\bphi, \bht] = -\bphi  \cdot (\cL_{\bht} \bm\bpi).$$
Finally we use the definition of $\kappa_{\t} = \bar{\gamma}_{\bht} - \bar\d_{\bht} - \bphi\cdot \bj$ to rewrite the previous equation as
\be
-\bphi  \cdot (\cL_{\bht}+\theta_{\bht}) \bm\bpi
+ \bphi\cdot \left( \brd \bar{\gamma}_{\bht} - \brd\cdot \bm{\widehat{\Theta}}_{\bot}\right)
   -\theta_{\bot}\cL_{\bphi}\phi  + T_{\bphi \bot}
       + \brd \left( \bphi (\theta_{\bot} - \kappa_{\t}) \right)
       +\frac12 \rd^{a}( \cL_{\bot} \varphi_{a})=0.\nn
\ee
Rearringing the terms this gives 
\be
\boxed{-\varphi^{a}\left( (\cL_{\bht}+\theta_{\bht}) \bm\bpi_{a}
+  \left( \brd \bar{\gamma}_{\bht} - \brd\cdot \bm{\widehat{\Theta}}_{\bot}\right)_{a}
   -\theta_{\bot}\rd_{a}\phi + (8\pi G) T_{a \bot} \right)
       + \rd^{a}\left( \varphi_{a} (\theta_{\bot} - \kappa_{\t})  
       +\frac12 ( \cL_{\bot} \varphi_{a}) \right)=0
}\label{mt}
\ee

\section{Form identities}\label{fid}
 We denote the differential on $M$ by $\rd$ and the interior product of a form $\omega$ by a vector $t$ by
 $\imath_{t}\omega$.
 The Lie derivative is given by 
 $\cL_{\t} = \rd \imath_{\t} +\imath_{t}\rd$.
 We introduce the following volume forms 
\bea
\epsilon &\equiv& \f1{4 !}\sqrt{|g|} \epsilon_{\a\b\gamma\delta} \rd x^{\a}\wedge  \rd x^{\b}\wedge\rd x^{\gamma}\wedge\rd x^{\delta} \nn\\
\epsilon_{\mu} &\equiv & \imath_{\partial_{\mu}} \epsilon =  \f1{3 !}\sqrt{|g|} \epsilon_{\mu\a\b\gamma} \rd x^{\a}\wedge  \rd x^{\b}\wedge\rd x^{\gamma}\nn\\
\epsilon_{\mu\nu} &\equiv &\imath_{\p_{\nu}} \imath_{\partial_{\mu}} \epsilon =  \f1{2 !}\sqrt{|g|} \epsilon_{\mu\nu\a\b} \rd x^{\a}\wedge  \rd x^{\b} \nn\\
\eea
They satisfy
\be
\rd (V^{\mu\nu} \epsilon_{\mu\nu}) = 2\N_{\nu}V^{[\mu\nu]} \epsilon_{\mu},\qquad
\rd (V^{\mu} \epsilon_{\mu}) = \N_{\mu}V^{\mu} \epsilon
\ee
and 
\be
\i_{\t}\epsilon = t^{\mu}\e_{\mu},\qquad \i_{\t} ( V^{\mu}\epsilon_{\mu})= V^{[\mu}t^{\nu]} \epsilon_{\mu\nu}.
\ee
From this we can show that their Lie derivative is given by
\be
\cL_{\t} \epsilon = (\N_{\a}t^{\a}) \epsilon,\qquad \cL_{\t} \epsilon_{\mu} = (\N_{\a}t^{\a}) \epsilon_{\mu}.
\ee

\section{ Boundary variation}\label{bdyvar}
Here we want to compute the following variational terms
entering the definition of the boundary symplectic potential \Ref{boundsymp}
\be
\bm\d \equiv s_{a}\delta n^{a} + s^{a}\delta n_{a},\qquad 
\bm{\bar\d} \equiv \bs_{a}\delta \bn^{a} + \bs^{a}\delta \bn_{a}.
\ee
since $(\bbn,\bbs)=\cosh\beta (\n,\s)+ \sinh\beta (\s,\n)$ it is clear that
\be\label{reld}
\bm{\bar{\d}} = \bm\d + 2 \d\beta.
\ee
One restricts for this computation  to variations that preserves the double foliation $\delta R =\delta T =cst$, this translates into
$\delta n_{a}\propto n_{a}$, $ \d \bs_{a}\propto \bs_{a}$ which implies that:
\be
s^{a}\d n_{a}=0= \bn^{a}\d \bs_{a}.
\ee
In order to evaluate this variation we need that 
\be
n_{a}\rd x^{a} =  -\rho \cosh \beta\, \rd T ,\qquad 
s_{a}\rd x^{a} =  \tau \rd R + \rho  \sinh \beta\, \rd T .
\ee
while the corresponding vectors are 
\bea
 \tau \partial_{\s} =  \partial_{R} - \psi^{A} \partial_{A}\equiv \partial_{\bm{\hat{r}}} \nn,\qquad
(\rho\tau \cosh \beta) \partial_{\n} =\tau \partial_{T}-\rho \sinh\beta \partial_{R}-(\varphi^{A}-\sinh\beta\psi^{A})\partial_{A},
\eea
for the slicing frame and 
These variations we can then evaluate directly
\bea
\bm\d &=& s_{a}\delta n^{a} = - n^{a}\d s_{a} = -  n^{R} \d s_{R}-  n^{T}\d s_{T} \nn\\
&=& \frac{\sinh \beta}{\tau\cosh \beta} \d \tau  - \frac{1}{\rho\cosh \beta} \d (\rho \sinh\beta)\nn\\
&=& -\d \beta + \tanh\beta\left(\frac{\delta\tau}{\tau}-\frac{\delta \rho}{\rho} \right)
\eea
Using the relationship \Ref{reld} between $\bm\d $ and $\bm{\bar\d}$ we get that
%and similarly
%For the screen frame  $(\bbn,\bbs)= \cosh \beta (\n,\s)  + \sinh \beta (\s,\n)$, we have:
%\be\label{bns}
%\bs_{a}\rd x^{a} =  \tau\cosh \beta \,\rd R,\qquad
%\bn_{a}\rd x^{a} =-\rho  \rd T + \tau \sinh \beta\, \rd R.
%\ee
%\bea
%\rho \partial_{\bbn} =\partial_{T}-\varphi^{A}\partial_{A}\equiv\partial_{\bht},\qquad
%(\rho \tau\cosh \beta) \partial_{\bbs} =  \rho \partial_{R}+\tau \sinh\beta \partial_{T} -( \psi^{A}+\sinh\beta\varphi^{A}) \partial_{A}\nn.
%\eea
%for the screen frame.
\be
\boxed{\bm{\bar{\d}}= \d \beta + \tanh\beta\left(\frac{\delta\tau}{\tau}-\frac{\delta \rho}{\rho} \right)
= {\tanh\beta}\d\ln \left(\frac{\tau}{\rho} \sinh\beta \right)}
\ee
%For completeness we also evaluate for foliation preserving variations
%\bea
%s^{a}\d s_{a}&=& s^{R} \d s_{R}= \frac{\d \tau}{\tau},\qquad
%n^{a}\d n_{a}= n^{T}\d n_{T}= -\frac{\d(\rho\cosh \beta)}{\rho\cosh \beta},
%\eea
%and similarly for the screen frame
%\bea
%\bs^{a}\d \bs_{a}&=&  \frac{\d (\tau\cosh\beta)}{\tau\cosh\beta},\qquad 
%\bn^{a}\d \bn_{a}=  -\frac{\d\rho}{\rho}.
%\eea

\section{Codazzi and Ricci equation}\label{Gauss}

We consider a spacetime $M$ endowed with a foliation.
The leaves  $\Sigma_t$ of this foliation are the level set of a given spacetime time function $T(x)$,
%$\Sigma_t = \{ x\in M/ T(x)=t\}$. 
We denote by $(g_{\mu\nu}, \N_{\nu})$ the spacetime metric and covariant derivative, 
we  also denote by  $(h_{\mu\nu}, D_{\nu})$ the metric and covariant derivative on $\Sigma_{t}$.
The unit normal to $\Sigma_t$ is denoted by $n$ and satisfies $n\cdot n=-1$ where dot means a contraction $n^\mu n_\mu$.
We can therefore relate the  metric and connection  on the slices to the spacetime ones by the use of the orthonormal projector
\be
h_{\mu}{}^{\nu} = g_{\mu}{}^{\nu} + n_{\mu} n^{\nu}, \qquad h_{\mu}{}^{\alpha} \nabla_{\alpha} v_{\nu} =
D_{\mu}v_{\nu}
\ee
for a vector $v$ tangent to $\Sigma_{t}$.
The time evolution is characterised by a time flow vector $\t=t^{\mu}\partial_{\mu}$ which can be decomposed in terms of a lapse and a shift 
\be
\t= N \n + \bm{M}.
\ee
The characteristic property of this vector is  that the Lie derivative along $t$ of any vector tangent to $\Sigma_{t}$ is still tangent to $\Sigma_{t}$. This means that 
\be
h_{\alpha}{}^{\mu}\cL_{\t} n_{\mu} =0.
\ee
and it implies  that the acceleration $a_{\mu}$ of fiducial observers static with respect to the foliation and whose velocity is given by $n^{\mu}$ is a vector tangent to $\Sigma_{t}$ given by the space derivative of the lapse function:
\be
a_{\mu}\equiv \nabla_{\n} n_{\mu} = \f{D_{\mu} N}{N}. 
\ee
If we denote by $T$ the time function characterising the foliation, its property are that

We denote by $g_{\mu\nu}$ the spacetime metric 
From the definition of the Riemann tensor we have that 
\bea
h_\alpha{}^{\alpha'} h_\beta{}^{\beta'} R_{\alpha'\mu\beta' \nu} t^\mu n^\nu 
&=& h_\alpha{}^{\alpha'} h_\beta{}^{\beta'} t^\mu ( \N_{\alpha'} \N_\mu n_{\beta'} - \N_\mu \N_{\alpha'} n_{\beta'}) \nn\\
&=&  h_\beta{}^{\beta'}  D_\alpha \N_{\t} n_{\beta'} -(h_\alpha{}^{\alpha'}\N_{\alpha'} t^\mu) (h_\beta{}^{\beta'}\N_\mu n_{\beta'})
- h_\alpha{}^{\alpha'} h_\beta{}^{\beta'} \N_{\t}( D_{\alpha'} n_{\beta'}-n_{\alpha'} a_{\beta'}) \nn
%\nn \\
%&=&    D_\alpha \N_{\t} n_{\beta} -(h_\alpha{}^{\alpha'}\N_{\alpha'} t^\mu)h_{\mu}{}^{\nu} (h_\beta{}^{\beta'}\N_\nu n_{\beta'})
%- h_\alpha{}^{\alpha'} h_\beta{}^{\beta'} \N_{\t}( D_{\alpha'} n_{\beta'}-n_{\alpha'} a_{\beta'}) \nn \\
%&=& h_\alpha{}^{\alpha'} h_\beta{}^{\beta'} \N_{\t} K_{\alpha' \beta'}+ K_{\mu\beta} D_\alpha t^\mu
%+ D_{\alpha} A_{\t \beta} \nn \\
%&=& 
\eea
In order to continue we introduce the extrinsic tensor 
$$ K_{\alpha\beta}\equiv D_{\alpha}n_{\beta}.$$ 
and 
the notation $$\boxed{D_{\alpha} t_{\beta} \equiv -N K_{\alpha \beta} + D_{\alpha}M_\beta.}$$
the symmetrisation of this tensor is the time derivative of the metric
\be
\cL_{\t} h_{\alpha\beta} = D_{\alpha} t_{\beta}+ D_{\beta} t_{\alpha},
\ee
We also introduce the important notion of the acceleration of the $\t$ observers defined by
\be
\boxed{a_{\t}^{\mu}\equiv \nabla_{\t}n^{\mu} = N a^{\mu} - K^{\mu\nu}M_{\nu}.}
\ee
Since the time flow preserve the foliation we have  
$ h_{\mu}{}^{\alpha}\cL_{\t}n_{\alpha} =\nabla_{\t}n_{\mu} + \n\cdot D_{\mu} \t =0 $, so we can equivalently write the $\t$-observer acceleration
as 
\be
a_{\t}^{\mu}= -  \n\cdot D^{\mu} \t.
\ee
This means that $ h_\alpha{}^{\alpha'} \N_{\alpha'} t^\mu = D_{\alpha}t^{\mu} +  a_{\t\alpha} n^{\mu},$
thus  we get 
\bea
h_\alpha{}^{\alpha'} h_\beta{}^{\beta'} R_{\alpha'\t\beta' \n}
&=&- h_\alpha{}^{\alpha'} h_\beta{}^{\beta'} \N_{\t} K_{\alpha' \beta'}- K_{\mu\beta} D_\alpha t^\mu
+ D_{\alpha} a_{\t \beta} \nn 
%\\
%&=& \cL_{t}K_{\alpha\beta} - K_{\alpha \mu} D_{\beta}t^{\mu} +  D_{\alpha} A_{\t \beta}
\eea
Using the definition of the Lie derivative  and the antisymetry of $R$ in the last two indices, we can write this equation as 
\be
\boxed{ h_\alpha{}^{\alpha'}  R_{\alpha'\t\beta \n} =  -\cL_{t}K_{\alpha\beta} + K_{\alpha \mu} D_{\beta}t^{\mu} +  D_{\alpha} a_{\t \beta}}
\ee
This is the combination of Ricci and Codazzi equation.
Taking the trace of this identity we obtain
\bea\label{Rictr}
\boxed{R_{\t\n}
= -\cL_{\t}K - K_{\alpha \beta} D^{\alpha}t^{\beta} +  D_{\alpha} \N_{\t}n^{\alpha}}
%&=& h^{\alpha\beta}\cL_{t}K_{\alpha\beta} - K_{\alpha\beta} D^{\alpha}t^{\beta} +  D_{\alpha} a_{\t}^{\alpha} \\
\eea
%we also have 
%\bea\label{Rictr}
%\boxed{2G_{\n\n}
%= R(h)- K: K + K^{2}}
%%&=& h^{\alpha\beta}\cL_{t}K_{\alpha\beta} - nK_{\alpha\beta} D^{\alpha}t^{\beta} +  D_{\alpha} a_{\t}^{\alpha} \\
%\eea
%and 
%\bea\label{Rictr}
% \boxed{ R(g)
%= 2\cL_{\n}K + K:K+K^{2} +R(h) -  2D_{\alpha} a_{\n}^{\alpha}}
%%&=& h^{\alpha\beta}\cL_{t}K_{\alpha\beta} - K_{\alpha\beta} D^{\alpha}t^{\beta} +  D_{\alpha} a_{\t}^{\alpha} \\
%\eea

\section{More on acceleration}\label{acc-sec}
Let us recall the definition of the accelerations in section \ref{int-ext}.
We have the normal accelerations 
\be
\gamma_{\n}\equiv \s\N_{\n}\n,\qquad \gamma_{\s}\equiv \s\N_{\s}\n.
\ee
And the tangential accelerations
\be
{\bm a}_{\n}=\N_{\n}\n,\quad
\bm{a}_{\s} = \N_{\s}\n,\quad
\bm{b}+\frac{\bm{\omega}}2 = \N_{\n}\s,\quad
\bm{b}-\frac{\bm{\omega}}2 = \N_{\s}\n.
\ee
plus the normal one form $\pi_{a}= q_{a}{}^{b}\s\N_{b}\n$.
These accelerations  appear in the decomposition of the differential of $\n,\s$:
\bea
\bm\rd \n &=& \gamma_{\n} \s \wedge \n + {\bm a}_{\n}\wedge \n + \left(\frac{\bm{\omega}}2+\bm \pi -\bm b\right)\wedge \s, \\
\bm\rd \s &=& \gamma_{\s} \s\wedge \n + \bm{a}_{\s}\wedge \s - \left(\frac{\bm{\omega}}2+\bm \pi +\bm b\right)\wedge \n.
 \eea
The space-time metric can be written in two different ways according to the time slicing or screen slicing:
\bea
\rd s^{2} &=& -\rho^{2} \cosh^{2}\beta \rd T^{2} + \left( \tau \rd R + \rho  \sinh \beta \rd T    \right)^{2}
+ \left(\rd \bm{x} + \bm\varphi \rd T + \bm\psi \rd R \right)\dd \bm q \dd
\left(\rd \bm{x} + \bm\varphi \rd T + \bm\psi \rd R \right) \nn \\
&=&  \tau^{2} \cosh^{2}\beta \rd R^{2} - \left( \rho \rd T - \rho  \sinh \beta \rd R    \right)^{2}
+ \left(\rd \bm{x} + \bm\varphi \rd T + \bm\psi \rd R \right)\dd \bm q \dd
\left(\rd \bm{x} + \bm\varphi \rd T + \bm\psi \rd R \right) \nn
\eea
Lets start with the time slicing frame $(\n,\s)$.
In this coordinates the orthonormal slice frame is given by
\be\label{ns}
n_{a}\rd x^{a} =  -\rho \cosh \beta\, \rd T ,\qquad 
s_{a}\rd x^{a} =  \tau \rd R + \rho  \sinh \beta\, \rd T .
\ee
For the screen frame  $(\bbn,\bbs)= \cosh \beta (\n,\s)  + \sinh \beta (\s,\n)$, we have:
\be\label{bns}
\bs_{a}\rd x^{a} =  \tau\cosh \beta \,\rd R,\qquad
\bn_{a}\rd x^{a} =-\rho  \rd T + \tau \sinh \beta\, \rd R.
\ee
while the corresponding vectors are 
\bea
 \tau \partial_{\s} =  \partial_{R} - \psi^{A} \partial_{A}\equiv \partial_{\bm{\hat{r}}} \nn,\qquad
(\rho\tau \cosh \beta) \partial_{\n} =\tau \partial_{T}-\rho \sinh\beta \partial_{R}-(\varphi^{A}-\sinh\beta\psi^{A})\partial_{A},
\eea
for the slicing frame and 
\bea
\rho \partial_{\bbn} =\partial_{T}-\varphi^{A}\partial_{A}\equiv\partial_{\bht},\qquad
(\rho \tau\cosh \beta) \partial_{\bbs} =  \rho \partial_{R}+\tau \sinh\beta \partial_{T} -( \psi^{A}+\sinh\beta\varphi^{A}) \partial_{A}\nn.
\eea
for the screen frame.
By taking the differential of \Ref{ns} formula and using that 
$$\rd T = -\frac{\n}{\rho\cosh\beta},\qquad  \rd R = \frac{1}{\tau}(\s +\tanh \beta \n) $$
we can express the accelerations in term of the foliation scalars: One obtains
\bea
\bm{ \rd n} &=&
\frac{\partial_{\bhr}(\rho \cosh \beta)}{\rho\tau \cosh \beta} \s \wedge \n 
+ \frac{\bm \rd (\rho \cosh \beta)}{\rho\cosh \beta}  \wedge \n \\
\bm{ \rd s} &=& \left(\frac{\partial_{\bht}\tau- \partial_{\bhr}(\rho \sinh \beta)}{\rho\tau \cosh \beta}\right) \s \wedge \n \nn
+ \frac{\bm\rd \tau}{\tau} \wedge \s 
- \tanh\beta  \bm\rd \ln \left(\frac{\rho \sinh \beta}{\tau }\right) \wedge \n\nn
\eea
from this we conclude that 
\be
\gamma_{\n} = \frac{\partial_{\bhr}(\rho \cosh \beta)}{\rho\tau \cosh \beta},\qquad
\gamma_{\s} =\frac{\partial_{\bht}\tau - \partial_{\bhr}(\rho \sinh \beta)}{\rho\tau \cosh \beta}
\ee
while
\be
{\bm a}_{\n}
= \frac{\bm\rd (\rho \cosh \beta)}{\rho \cosh \beta},
\qquad 
 {\bm a}_{\s}
= \frac{\bm\rd \tau}{\tau },
 \ee
 and 
\be
\N_{\s}\n =\bm \pi ,
\qquad \N_{\n}\s =- \bm \pi + \tanh \beta \left(\frac{\bm\rd \rho}{\rho}-\frac{\bm\rd \tau}{\tau}   \right) +  \bm\rd \beta .
 \ee
 Thus $\bm\omega=-2\bm\pi + 2\bm b $ with
 \be\boxed{
 2\bm{b}=  \tanh \beta \left(\frac{\bm\rd \rho}{\rho}-\frac{\bm\rd \tau}{\tau}   \right) +  \bm\rd \beta .}
 \ee
We can perform a similar computation in the screen frame
\bea
\bm{ \rd \bbs} &=&
\frac{\partial_{\bht}(\tau \cosh \beta)}{\rho\tau \cosh \beta} \bbs \wedge \bbn 
+ \frac{\bm \rd (\tau \cosh \beta)}{\tau\cosh \beta}  \wedge \bbs \\
\bm{ \rd \bbn} &=& \left(\frac{\partial_{\bhr}\rho+\partial_{\bht}(\tau \sinh \beta)}{\rho\tau \cosh \beta}\right) \bbs \wedge \bbn \nn
+ \frac{\bm\rd \rho}{\rho} \wedge \bbn 
+ \tanh\beta  \bm\rd \ln \left(\frac{\tau \sinh \beta}{\rho }\right) \wedge \n\nn
\eea
from this we conclude that 
\be
\bar\gamma_{\bbs} = \frac{\partial_{\bht}(\tau \cosh \eta)}{\rho\tau \cosh \eta},\qquad
\bar\gamma_{\bbn} =\frac{\partial_{\bhr}\rho + \partial_{\bht}(\tau \sinh \eta)}{\rho\tau \cosh \eta}
\ee
while
\be
{\bm {\bar{a}}}_{\bbn}
= \frac{\bm\rd \rho }{\rho },
\qquad \bar{\bm a}_{\bbs}
= \frac{\bm\rd (\tau\cosh\beta)}{\tau \cosh\beta}=\frac{\brd \tau}{\tau} +\tanh\beta \brd \beta,
 \ee
 and 
 \be
 \N_{\bbn}\bbs =-{\bm \bpi},
\qquad \N_{\bbn}\bbs = \bm\bpi - \tanh \beta \left(\frac{\bm\rd \tau}{\tau}-\frac{\bm\rd \rho}{\rho}   \right) -  \bm\rd \beta.
 \ee
 Thus 
 \be
 -2\bm{\bar b}=  \bm\rd \beta +  \tanh \eta \left(\frac{\bm\rd \tau}{\tau}-\frac{\bm\rd \rho}{\rho}   \right) .
 \ee
  Note that we can rewrite the radial acceleration coefficient which enters the definition of the surface tension as 
 \be\label{barg}
{\bar\gamma_{\bht} =\partial_{\bht}\beta + \frac{\partial_{\bhr}\rho + \sinh \beta \partial_{\bht}\tau }{\tau \cosh \beta}
 }
 \ee
 Using the fact that $ \frac{\s}{\cosh\beta} =  \bbs- \tanh \beta \bbn$ we can rewrite this expression as 
 \be\label{barg2}
 \bar\gamma_{\bht} =\partial_{\bht}\beta 
 + \tanh\beta\left( \frac{ \partial_{\bht}\tau }{\tau } -\frac{\partial_{\bht}\rho}{\rho}\right)
+ \frac{\partial_{\bot}\rho}{\rho}.
 \ee
 with $\bht=\rho\bbn$ and $\bot=\rho\bbs$.
 Note that from sec.\ref{sec-manyacc} and  eq.\ref{acc3}
  we have that
$\kappa_{\bht}= \frac12(\bar{\gamma}_{\bht}+ \N_{\bbs}\rho)$ is the average of the screen acceleration and the Newtonian acceleration.  On the other hand from the original definition we have that 
$\kappa_{\bht}= \bar{\gamma}_{\bht}- \frac12 \bar{\delta}_{\bht}$.
Equating the two expressions therefore gives that $\bar{\delta}_{\bht} = (\bar{\gamma}_{\bht} - \partial_{\bbs}\rho)$ is the difference between the radial and Newtonian acceleration, which is then equal to
\be
\boxed{
\bar{\delta}_{\bht} =
\partial_{\bht}\beta + \tanh \beta \left(\frac{\partial_{\bht}\tau}{\tau} -\frac{\partial_{\bht}{\rho}}{\rho}\right).
}
\ee
In summary this gives 
 \be
 \kappa_{\bht} =
 \N_{\bbs}\rho +\frac12\left(
\partial_{\bht}\beta + \tanh \beta \left(\frac{\partial_{\bht}\tau}{\tau} -\frac{\partial_{\bht}{\rho}}{\rho}\right) \right)
 \ee
 Now we also have that 
 \be
 \bm\kappa= -\bm\bpi + \frac12\left(\bm\rd \beta +  \tanh \eta \left(\frac{\bm\rd \tau}{\tau}-\frac{\bm\rd \rho}{\rho}   \right)\right)
 \ee
 so in total we get 
 \be
 \boxed{
 \kappa_{\t} =
 \N_{\bot}\phi - \bphi\dd \bm\bpi +\frac12\left(
\partial_{\t}\beta + \tanh \beta \left(\frac{\partial_{\t}\tau}{\tau} -\frac{\partial_{\t}{\rho}}{\rho}\right) \right)
 }
 \ee
 and 
 \be
 \boxed{
 \gamma_{\t} =
 \N_{\bot}\phi - \bphi\dd \bm\bpi +
\partial_{\t}\beta + \tanh \beta \left(\frac{\partial_{\t}\tau}{\tau} -\frac{\partial_{\t}{\rho}}{\rho}\right) 
 }
 \ee

\end{document}